\documentclass[11pt]{amsart}
\usepackage{amsaddr}
\usepackage{graphicx,amssymb,amsmath,amsthm}

\usepackage[utf8]{inputenc}
\usepackage{epstopdf}
\usepackage{bm}
\usepackage[numbers]{natbib}
\usepackage{color}
\usepackage{array}

\usepackage[margin=1.25in]{geometry}
\newcolumntype{L}{>{$}l<{$}} 
\newcolumntype{C}{>{$}c<{$}} 

\usepackage{bm,graphicx,textcomp,bbm} 

\usepackage{booktabs}

\usepackage{subcaption}
\newcommand{\R}{\mathbb{R}}
\newcommand*{\Id}{\operatorname{Id}}

\newcommand{\RMSE}{\mathcal{E}}
\newcommand{\RMSS}{\mathcal{S}}
\newcommand{\Dens}{{\bar{\mathcal{D}}}_{ens}}
\newcommand{\rrm}{}

\def\H{{\bf{H}}}

\def\y{{\bf{y}}}
\def\z{{\bf{z}}}

\def\Tr{{\bf{Tr}}}

\def\Z{{\bf{Z}}}
\def\P{{\bf{P}}}
\def\e{{\bf{e}}}

\def\xobs{{\bf{\y_{\rm{o}}}}}
\def\zobs{{\bf{\y_{\rm{o}}}}}

\def\robs{{\bf{r}}_{\rm{o}}}
\def\Robs{{\bf{R}}_{\rm{o}}}

\def\KR{{\bf{K}}}

\usepackage{amsfonts}

\title{Stochastically perturbed bred vectors in multi-scale systems}
\author{Brent Giggins and Georg A. Gottwald}
\address{School of Mathematics and Statistics, University of Sydney, NSW 2006, Australia}
\email[B. Giggins and G. A. Gottwald]{brent.giggins@sydney.edu.au {\rmfamily and} georg.gottwald@sydney.edu.au}

\begin{document}

	\begin{abstract}
		The breeding method is a computationally cheap way to generate flow-adapted ensembles to be used in probabilistic forecasts. Its main disadvantage is that the ensemble may lack diversity and collapse to a low-dimensional subspace. To still benefit from the breeding method's simplicity and its low computational cost, approaches are needed to increase the diversity of these bred vector (BV) ensembles. We present here such a method tailored for multi-scale systems. We describe how to judiciously introduce stochastic perturbations to the standard bred vectors leading to stochastically perturbed bred vectors. The increased diversity leads to a better forecast skill as measured by the RMS error, as well as to more reliable ensembles quantified by the error-spread relationship, the continuous ranked probability score and reliability diagrams. Our approach is dynamically informed and in effect generates random draws from the fast equilibrium measure conditioned on the slow variables. We illustrate the advantage of stochastically perturbed bred vectors over standard BVs in numerical simulations of a multi-scale Lorenz 96 model.
	\end{abstract}
	
	
	\maketitle
	
	
	\section{{\bf{Introduction}}}
	
	Weather and climate forecasting faces the problems that the underlying dynamics is inherently chaotic with often high sensitivities to small errors in the initial conditions \citep{Lorenz63} and involves coupled processes running on spatial scales from millimetres to thousands of kilometres, and temporal scales from seconds to millennia. This implies that in many situations generating a single forecast fails to provide sufficiently accurate information about a future state of the system and has to be viewed as only one particular realisation drawn from a high-dimensional probability distribution function. In chaotic dynamical systems probabilistic forecasts are more appropriate, and one seeks to estimate not only the expected state of the atmosphere but also some measure of the reliability of the forecast. A commonly used method to produce probabilistic forecasts is ensemble forecasting in which a Monte-Carlo estimate of the probability density function is obtained from multiple simulations, each starting from a different initial condition \citep{Epstein69,Leith74,LeutbecherPalmer08}. The high phase-space dimension of the atmosphere, however, would require large ensembles to estimate the full probability density function, which is out of reach given current computational resources. An attractive method to generate initial conditions for an ensemble forecast with low computational cost is the "breeding method" introduced by \cite{TothKalnay93,TothKalnay97}. The breeding method generates initial conditions which, rather than being random draws, encode information about locally fast growing modes. The rationale behind this is that the probability density function is supported in phase-space by regions which grew rapidly. It was used for more than a decade after its introduction in 1992 at the National Centers for Environmental Prediction (NCEP) for their operational 1-15 day ensemble forecasts, and has been widely used in atmosphere and climate probabilistic forecasts such as ENSO prediction \citep{CaiEtAl03,ChengEtAl10}, seasonal-to-interannual forecasting in coupled general circulation models (CGCMs) \citep{YangEtAl09} and forecasting Mars' weather and climate \citep{NewmanEtAl04,GreybushEtAl13}.\\
	
	In the breeding method a fiducial trajectory is propagated with the full nonlinear model starting from an (analysed) initial condition, along with an ensemble of nearby trajectories initialised from perturbed initial conditions, propagated under the same dynamics. To avoid saturation of instabilities, the perturbed trajectories are periodically rescaled to be of some finite-size distance $\delta$ away from the fiducial trajectory. 
	The difference at the time of rescaling is coined as the bred vector (BV). The breeding algorithm is conceptually related to the method for generating Lyapunov vectors, but in contrast to Lyapunov vectors, bred vectors are calculated by using the full nonlinear model. 
	Most importantly, Lyapunov vectors employ an infinitesimal perturbation size $\delta \to 0$, whereas BVs are generated using finite-size perturbations. This is motivated by the observation that typically the most unstable processes, i.e. those with the largest Lyapunov exponent, are small-scale processes, but become nonlinearly saturated at a much smaller level than slower growing large-scale instabilities. Hence, we can adjust the finite size of the perturbation to select the amplitude range of the instabilities to target specific growing modes. For example, BVs project onto baroclinic instabilities when the perturbation size is comparable to $1-10\%$ of the natural variability in the atmosphere \citep{TothKalnay97,CorazzaEtAl03}. Furthermore, in multi-scale systems that exhibit regimes, regime changes can be predicted for perturbation sizes in a certain range \citep{PenaEtAl04,NorwoodEtAl13}. We remark that in contrast to Lyapunov vectors which are mapped by the linear tangent dynamics onto each other, BVs with finite perturbation sizes are technically not vectors and there is no linear map relating them. We will adopt here the point of view that the object of interest are the perturbed states themselves, which serve as initial conditions, rather than the differences between them and a control run. We nevertheless keep with the convention of calling them vectors.\\
	
	It has long been noticed that bred vector ensembles lack diversity as they collapse onto a low dimensional subspace, and in the worst case most of the ensemble forecast variability is contained in a single BV \citep{WangBishop03,WeiEtAl06,Bowler06,Palmer18}. This effective reduction in ensemble size impedes their usage in sampling the forecast probability density function. For small perturbation sizes, bred vectors naturally align with the leading Lyapunov vector associated with the largest Lyapunov exponent leading to an effective ensemble dimension of one\footnote{This is strictly speaking only true locally; the sign of the disjoint local patterns is arbitrary \citep{TothKalnay97}.}. In a realistic operational forecasting situation uncertainty in the saturated sub-synoptic scales such as convective events may provide sufficient stochasticity to prevent bred vectors from collapsing into a single BV \citep{TothKalnay97}. Nevertheless, despite the presence of unresolved sub-synoptic perturbations in realistic forecast systems, BVs tend to be under-dispersive \citep{Palmer18}.\\
	
	Several mitigation strategies have been devised to increase the effective size of an ensemble of bred vectors. \cite{Annan04,KellerEtAl10} proposed to orthogonalise the BVs. \cite{WeiEtAl08} combined BVs with the ensemble transform Kalman filter \citep{BishopEtAl01,TippettEtAl03} to increase the diversity. \cite{FengEtAl14,FengEtAl16,FengEtAl18} introduced nonlinear local Lyapunov vectors as an orthogonalised modification to classical BVs. \cite{OKaneFrederiksen08} employed stochastic backscattering to increase the diversity. \cite{PrimoEtAl08,PazoEtAl11,PazoEtAl13} generated BVs employing the geometric rather than the Euclidean norm when rescaling. \cite{BalciEtAl12} proposed a different rescaling procedure based on the largest BV. \cite{GreybushEtAl13} added small random perturbations to the BVs at each rescaling period, an idea already proposed by \cite{KalnayTalk08}. Here we will introduce two new methods to stochastically modify bred vectors, targeted at multi-scale systems to increase the ensemble spread by increasing the diversity of perturbation patterns. We introduce stochastically perturbed bred vectors (SPBV) and random draw bred vectors (RDBV). Our method to generate SPBVs is dynamically informed and views BVs as providing initial conditions which are likely to be good candidates to be used as a Monte-Carlo estimate for the future probability density function. We exploit the fact that the joint probability measure in a slow-fast system can be approximated by a product measure comprised of the measure of the slow variables and the equilibrium measure of the fast variables conditioned on the slow variables. The collapse of BVs for small values of the perturbation size then implies that we have only one single random draw from the probability measure of the fast variables conditioned on the slow variables. 
	%
	We present a way to generate a diverse ensemble from this collapsed ensemble representing an ensemble drawn from this fast conditional probability density function by a cost-effective stochastic perturbation. We further consider random draws from the marginal fast equilibrium probability density function to generate random draw bred vectors (RDBV). In RDBVs, as opposed to SPBVs, the fast components are drawn independent of the current state of the slow components.  The stochastically modified bred vectors both introduce variance in the sub-synoptic scales, where our uncertainty about the current state is largest, consistent with an analysed state obtained from data assimilation. We will show that the sub-synoptic variance propagates into synoptic scales where most of the energy resides.\\
	
	We shall illustrate in numerical simulations of a multi-scale Lorenz-96 model (L96, \cite{Lorenz96}) several advantages of SPBVs and RDBVs over the standard BVs. The ensemble dimension of SPBVs and RDBVs is significantly increased, in particular for small but finite values of the perturbation size $\delta$. This increased diversity leads to much better forecasting skill when compared to the standard BVs, with RDBVs having lower forecast error than SPBVs. Furthermore, whereas the RMS forecast errors of standard BVs drop significantly when the perturbation size $\delta$ is increased from small values for which BVs align with the leading Lyapunov vector to values which correspond to the nonlinear regime, the RMS error varies smoothly for SPBVs and RDBVs. This has the advantage that the forecast skill is much less sensitive to the choice of the perturbation size which in practice should be chosen to be compatible with the analysis covariance provided by, for example, data assimilation \citep{TothKalnay97,WangBishop03,CorazzaEtAl03}. In probabilistic forecasts, the reliability of an ensemble is of great importance. An ensemble is called reliable if the associated forecast probability provides an unbiased estimate of the observed relative frequencies. 
Using error-spread relationships, the continuous ranked probability score and Talagrand diagrams we show that SPBV and RDBV ensembles are more reliable than the standard BVs.   
Bred vectors also have the desirable feature that they are dynamically consistent in the sense that their time evolution is close to the actual dynamics of the dynamical multi-scale system. Using the mean-variance diagrams for the logarithm of the bred vectors, which illustrate characteristic features of the temporal evolution of errors in chaotic dynamical systems, we show that SPBVs inherit the dynamical consistency whereas RDBVs, despite having better forecast skill, lack dynamical consistency.\\
	
The paper is organised as follows. In Section~\ref{sec.model} we introduce the multi-scale Lorenz-96 model \citep{Lorenz96}. Section~\ref{sec.BV} provides a brief introduction to bred vectors and the breeding method. In Section~\ref{sec.SPBV} we introduce our dynamically informed modified method of stochastically perturbed bred vectors and the dynamically inconsistent method of random draw bred vectors. We then continue to show the efficiency of our stochastically modified bred vectors in numerical simulations using the L96 model. Section~\ref{sec.numerics} presents numerical simulations illustrating the advantages of stochastically modified bred vectors over classical bred vectors. In Section~\ref{sec.Dens} we show how the ensemble dimension is increased, and in Sections~\ref{sec.skill}-\ref{sec.reliability} we present results on the forecast skill and the reliability of the ensemble, respectively. The dynamical features of bred vectors and their stochastically perturbed counterparts are investigated in Section~\ref{sec.evol}. Here we show how bred vectors project onto covariant Lyapunov vectors, and investigate the dynamical consistency of bred vectors by looking at their mean-variance of the logarithm (MVL) diagram. We conclude in Section~\ref{sec.summary} with a discussion and an outlook.
	
	
\section{{\bf{The Multi-Scale Lorenz-96 System}}}
	\label{sec.model}
	
	We consider the multi-scale Lorenz 96 system \citep{Lorenz96}, which was introduced as a caricature for the atmosphere. The model describes $K$ slow variables $X_k$ which are each coupled to $J$ fast variables $Y_{j,k}$, governed by the following equations 
	\begin{align}
	\frac{d}{dt}X_k &= -X_{k-1}(X_{k-2}-X_{k+1})-X_k+F-\frac{hc}{b}\sum\limits_{j=1}^{J}Y_{j,k}, 
	\label{e.L96_X}\\
	\frac{d}{dt}Y_{j,k} &= -cbY_{j+1,k}(Y_{j+2,k}-Y_{j-1,k})-cY_{j,k}+\frac{hc}{b}X_k,  
	\label{e.L96_Y}
	\end{align}
	with cyclic boundary conditions $X_{k+K}=X_k$, $Y_{j,k+K}=Y_{j,k}$ and $Y_{j+J,k} = Y_{j,k+1}$, giving a total of $D=K(J+1)$ variables. The variables $X_k$ can be interpreted as large scale atmospheric fields arranged on a latitudinal circle, such as synoptic weather systems. Each of the $X_k$ variables is connected to $J$ small-scale variables $Y_{j,k}$ with smaller amplitude and frequency, modelling for example convective events. The coefficient $c$ signifies the time-scale separation, and the ratio of the amplitudes of the large-scale and the small-scale variables is controlled by $b$. The coupling strength is given by the parameter $h$. The uncoupled dynamics of both the large-scale and the small-scale variables is given by nonlinear transport and linear damping; the large-scales are subjected to external forcing $F$.  
	
	\begin{table}[h]
		\begin{center}
			\begin{tabular}{c c c}
				\toprule
				\textbf{Parameter} & \textbf{\ Description \ } & \textbf{Value}\\
				\midrule
				$K$ & \ number slow variables \ & 12 \\
				$J$ & \ number fast variables per slow variable \ & 24 \\
				$c$ & \ time-scale ratio \ & 10 \\
				$b$ & \ amplitude ratio \ & 10  \\
				$F$ & \ forcing \ & 20 \\
				$h$ & \ coupling constant \ &  1\\
				\bottomrule
			\end{tabular}		
			\caption {Parameters used for the multi-scale L96 System (\ref{e.L96_X})-(\ref{e.L96_Y}).}
			\label{tab:L96} 
		\end{center}
	\end{table}
	
	We select parameter values, listed in Table \ref{tab:L96}, leading to chaotic behaviour. The choice $c=b=10$ implies that the variables  $Y_{j,k}$ fluctuate with a $10$ times higher frequency and with an approximately $10$ times smaller amplitude when compared to the $X_k$. We also set the coupling constant $h=1$, corresponding to strong coupling where the dynamics is driven by the fast sub-system \citep{HerreraEtAl10}. For the parameters given in Table~\ref{tab:L96} the climatic variance is estimated as $\sigma^2_{X, \rm{clim}}=31.62$ for the slow variables and as $\sigma^2_{Y, \rm{clim}} = 0.1061$ for the fast variables. We also estimate the decorrelation ($e$-folding) time of the slow variables to be $\tau_{X,c}=0.1705$ and of the fast variables is  $\tau_{Y,c}=0.0345$. The maximal Lyapunov exponent is measured as $\lambda_{\rm{max}}=18.29$. \\
	
\begin{figure}[h]
		\centering
		\begin{subfigure}{.5\textwidth}
			\centering
			\caption{}
			\includegraphics[width=0.95\linewidth]{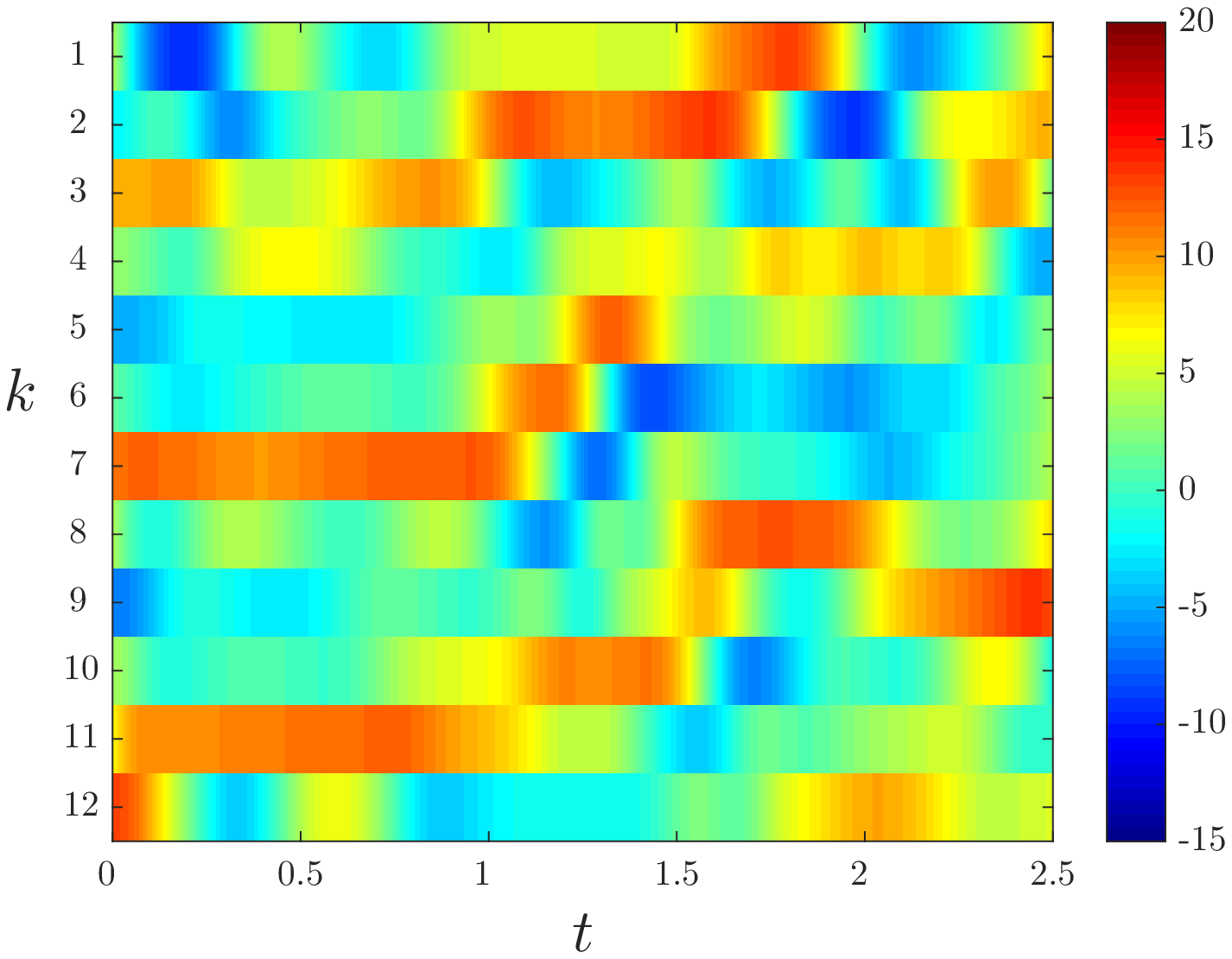}		
			\label{fig:L96heatmapslow}
		\end{subfigure}%
		\begin{subfigure}{.5\textwidth}
			\centering
			\caption{}
			\includegraphics[width=0.95\linewidth]{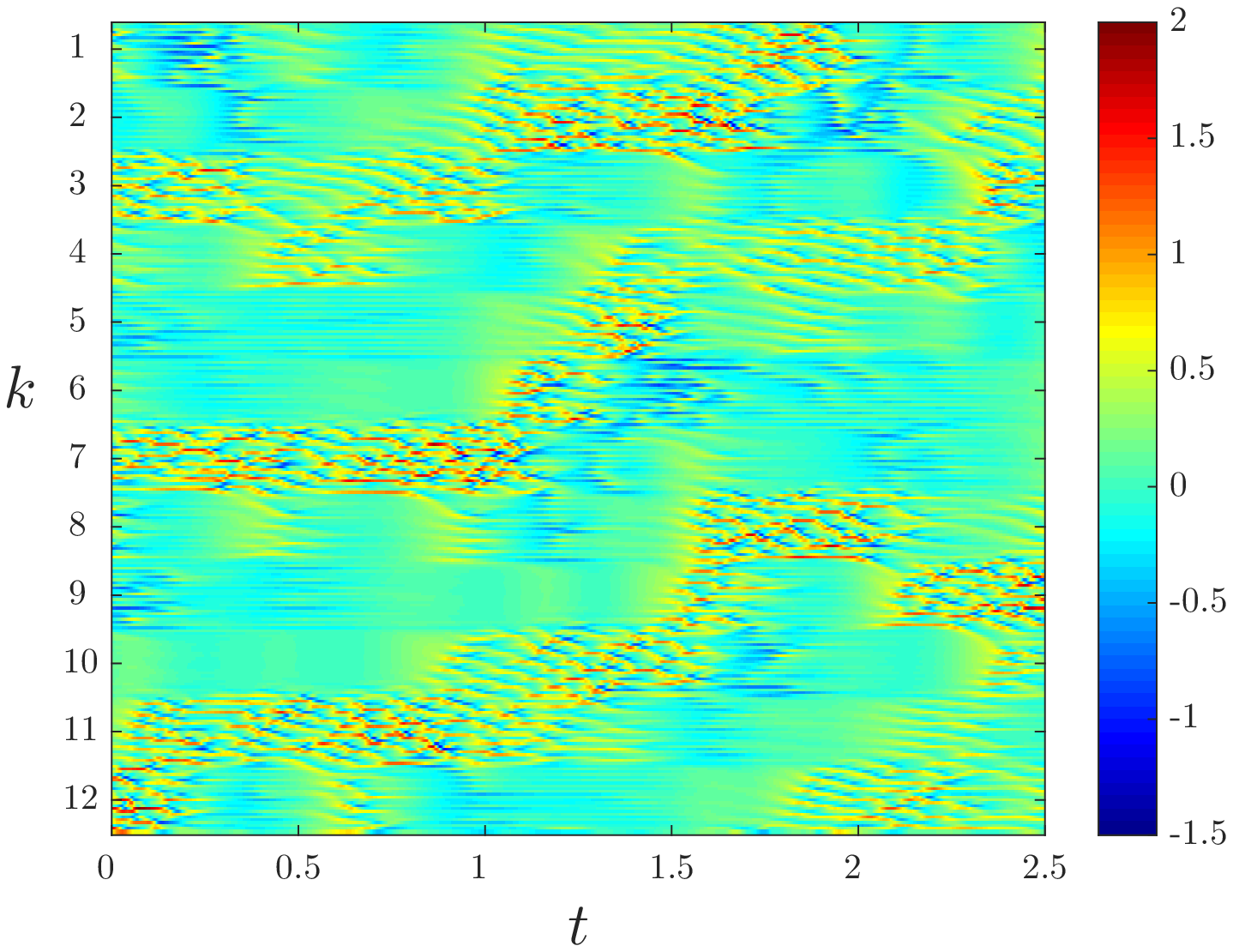}		
			\label{fig:L96heatmapfast}
		\end{subfigure}
		\caption{Spatio-temporal dynamics of the multi-scale L96 model (\ref{e.L96_X})-(\ref{e.L96_Y}) for (a) large-scale variables $X$ and (b) small-scale variables $Y$.}
		\label{fig:L96heatmap}
	\end{figure}

\begin{figure}[h]
		\centering
		\includegraphics[width=0.6\linewidth]{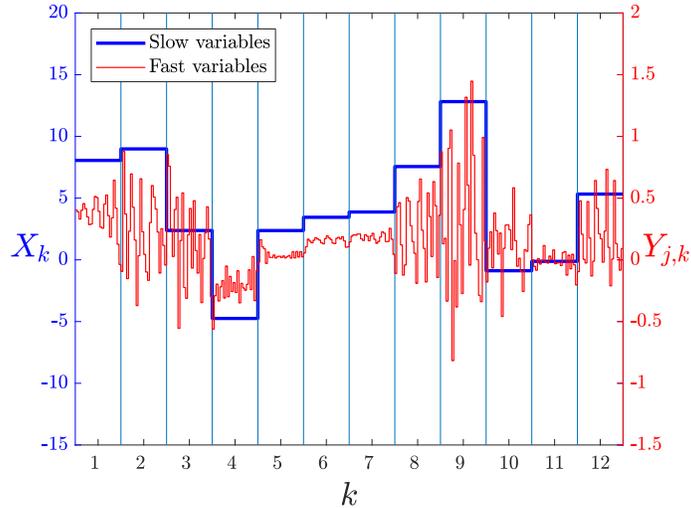}
		\caption{Snapshot of the large-scale variables $X_k$ and the small-scale variables $Y_{k,j}$ for the multi-scale L96 model (\ref{e.L96_X})-(\ref{e.L96_Y}) (corresponding to Figure~\ref{fig:L96heatmap} at $t=2.5$). Shown are the $K=12$ slow variables $X_k$ and for each slow sector the $J=24$ fast variables $Y_{k,j}$. }
		\label{fig:L96timesnapshot}
	\end{figure}

	To numerically simulate the multi-scale L96 system we employ a fourth-order Runge-Kutta method with a fixed time step $dt = 0.0005$. In our simulations we employ an initial transient time of $250$ time units to assure that the dynamics will have settled on the attractor. A typical time series of the slow and fast variables is shown in Figure~\ref{fig:L96heatmap}. It is seen that the small-scale activity is regionally localised and is correlated to excited large-scale variables $X_k$. Figure~\ref{fig:L96timesnapshot} provides a snapshot of the state of the system at time $t=2.5$.
	
\section{{\bf{Bred Vectors and the breeding method}}}
	\label{sec.BV}
	
\begin{figure}[h]
		\centering
		\includegraphics[width=0.5\linewidth]{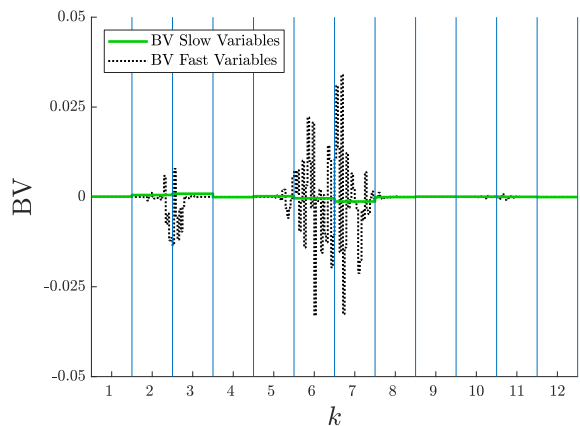}
		\caption{Bred vector ${\rm{BV}}$ for the multi-scale L96 model (\ref{e.L96_X})-(\ref{e.L96_Y}) with perturbation size $\delta=0.1$. Shown are the $K=12$ slow components and, for each slow sector, the $J=24$ fast components.}
		\label{fig:bvsnapshot}
	\end{figure}

	We first briefly describe the standard breeding method developed by \cite{TothKalnay93,TothKalnay97} before introducing our stochastically modified bred vectors. BV's are finite-size, periodically rescaled perturbations generated from the full non-linear dynamics of the system. Given a control trajectory $\boldsymbol{z}_c(t_i)$ at some time $t_i$, we define a perturbed initial condition 
	\begin{align*}
	\boldsymbol{z}_p(t_i) = \boldsymbol{z}_c(t_i) + \delta \frac{\boldsymbol{p}}{\| \boldsymbol{p} \|},
	\end{align*}
	where $\boldsymbol{p}$ is an initial arbitrary random perturbation and $\delta$ is the size of the perturbation. In realistic applications, control trajectories $\boldsymbol{z}_p(t_i)$ are seeded from an analysed state. The control and the perturbed initial condition are simultaneously evolved using the full non-linear dynamics for some integration time window $T$ until $t=t_{i+1}=t_i+T$. At the end of the integration window the difference between the two trajectories is calculated
	\begin{align*}
	\Delta \boldsymbol{z}(t_{i+1}) = \boldsymbol{z}_p(t_{i+1}) - \boldsymbol{z}_c(t_{i+1})
	\end{align*} 
	and the bred vector is defined as the difference rescaled to size $\delta$ with
	\begin{align*}
	\boldsymbol{b}(t_{i+1}) = \delta \frac{\Delta \boldsymbol{z}(t_{i+1})}{\| \Delta \boldsymbol{z}(t_{i+1}) \|}.
	\end{align*}
	The perturbation $\boldsymbol{b}(t_{i+1})$ is then used to redefine the perturbed trajectory $\boldsymbol{z}_p(t_{i+1}) = \boldsymbol{z}_c(t_{i+1}) + \boldsymbol{b}(t_{i+1})$ at the start of the next breeding cycle. For the L96 system we employ a breeding cycle length of $T=0.005$ time units. This process of breeding is repeated for several cycles until the growth rate of perturbations saturates and until the perturbations converge in the sense that at time $t_i$ the BVs span the same space as BVs obtained if the breeding cycle was initialised further in the past. These converged BVs are then employed for ensemble forecasts. In our simulations we employ a spin-up time for the BVs of $25$ time units (which amounts to $5000$ breeding cycles). An ensemble of $N+1$ initial conditions from which to start an ensemble forecast is then provided by adding $N$ separate BVs (that have each started from different initial perturbations $\boldsymbol{p}$) to the control and by the control itself. Figure~\ref{fig:bvsnapshot} shows a snapshot of a typical BV for $\delta=0.1$, revealing their localised character. The slow components of the bred vector are small and of order of magnitude of $10^{-3}$ at some active sites (here at the sites $k=2,3$ and at sites $k=6,7$) and otherwise are even smaller with amplitudes of the order of $10^{-5}$. The fast components of the BV are generally localised to those regions which correspond to the small but non-zero activity of the most dominant slow components. \\
	
	A forecast ensemble should consist of a diverse set of initial conditions that project onto likely areas of error growth in phase space. The performance of a bred vector ensemble depends on the perturbation size $\delta$. Ideally, the perturbation size should correspond to the analysis error \citep{TothKalnay97}. In practice, however, the perturbation size may have to be inflated to ensure that the evolving perturbations acquire sufficient spread at the desired lead time (see, for example, \cite{TothKalnay97,MagnussonEtAl08}). 
	If the perturbation size is chosen too large, the RMS error of the forecast will deteriorate; on the other hand, for small values of the perturbation size the spread of the ensemble may remain too small and not overlap with the truth, leading to poor forecasts. Furthermore, alignment with the leading Lyapunov vector (LLV) may lead to an ensemble collapse. Indeed, in the limit $\delta \to 0$ the perturbations are infinitesimal and after multiple spinup cycles the BV aligns with the LLV, with an average growth rate equal to the maximal Lyapunov exponent. For the BV depicted in Figure~\ref{fig:bvsnapshot} with $\delta=0.1$, we observe that an ensemble of $N=20$ BVs, which were initialised with different random perturbations, all collapse and are indistinguishable by eye from the one depicted in Figure~\ref{fig:bvsnapshot}.  
The lack of diversity of an ensemble of bred vectors and the collapse to the LLV for perturbation sizes corresponding to the analysis error of the day constitutes a major draw back of bred vectors.    
In the following we will devise a method how to generate a diverse ensemble of BVs.
	
\section{{\bf{Stochastically Perturbed Bred Vectors}}}
	\label{sec.SPBV}
\begin{figure}[h]
		\centering
		\begin{subfigure}{.49\textwidth}
			\centering
			\caption{}
			\includegraphics[width=0.98\linewidth]{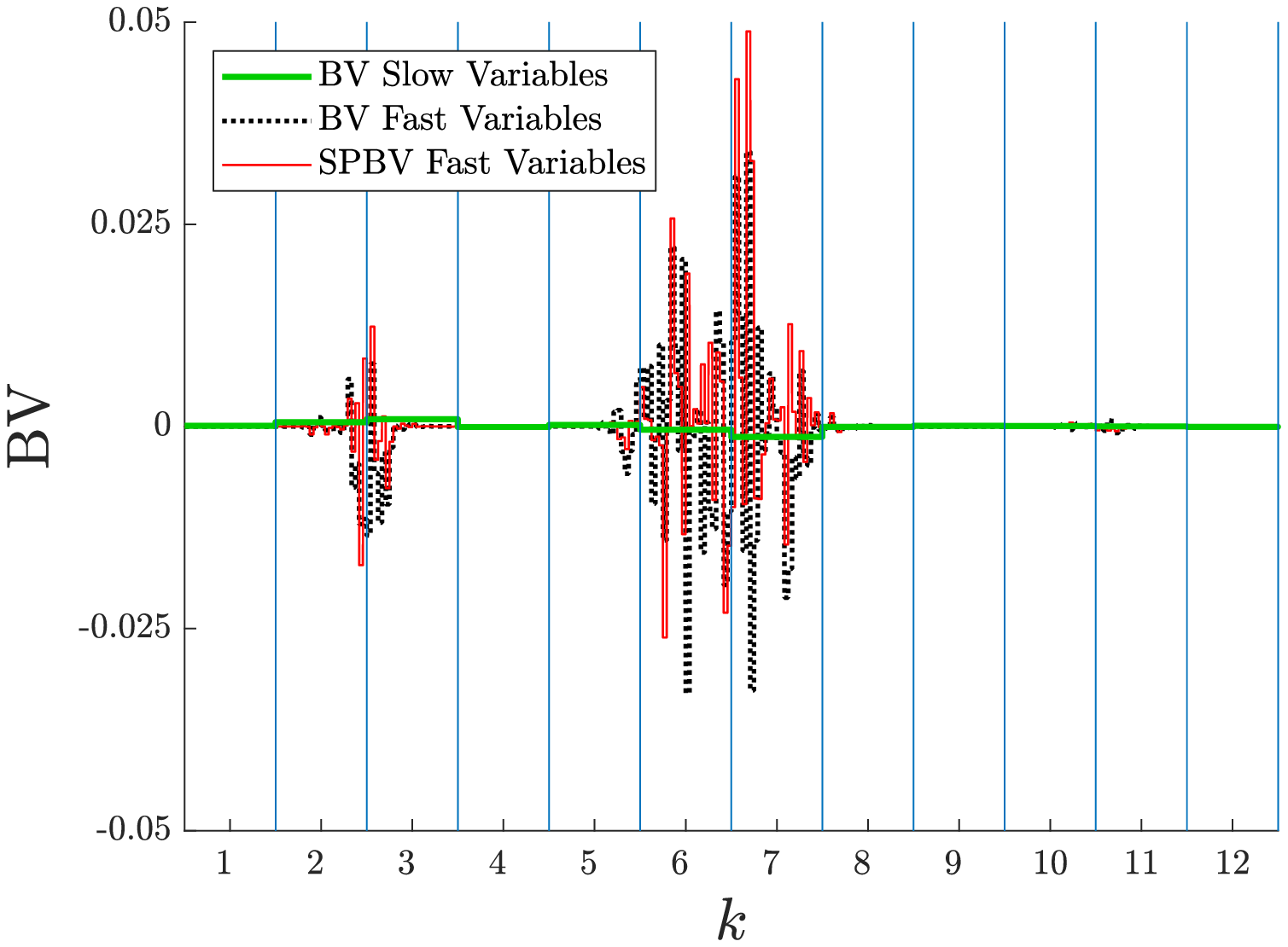}		
			\label{fig:spbvsnapshot}
		\end{subfigure}
		\begin{subfigure}{.49\textwidth}
			\centering
			\caption{}
			\includegraphics[width=0.98\linewidth]{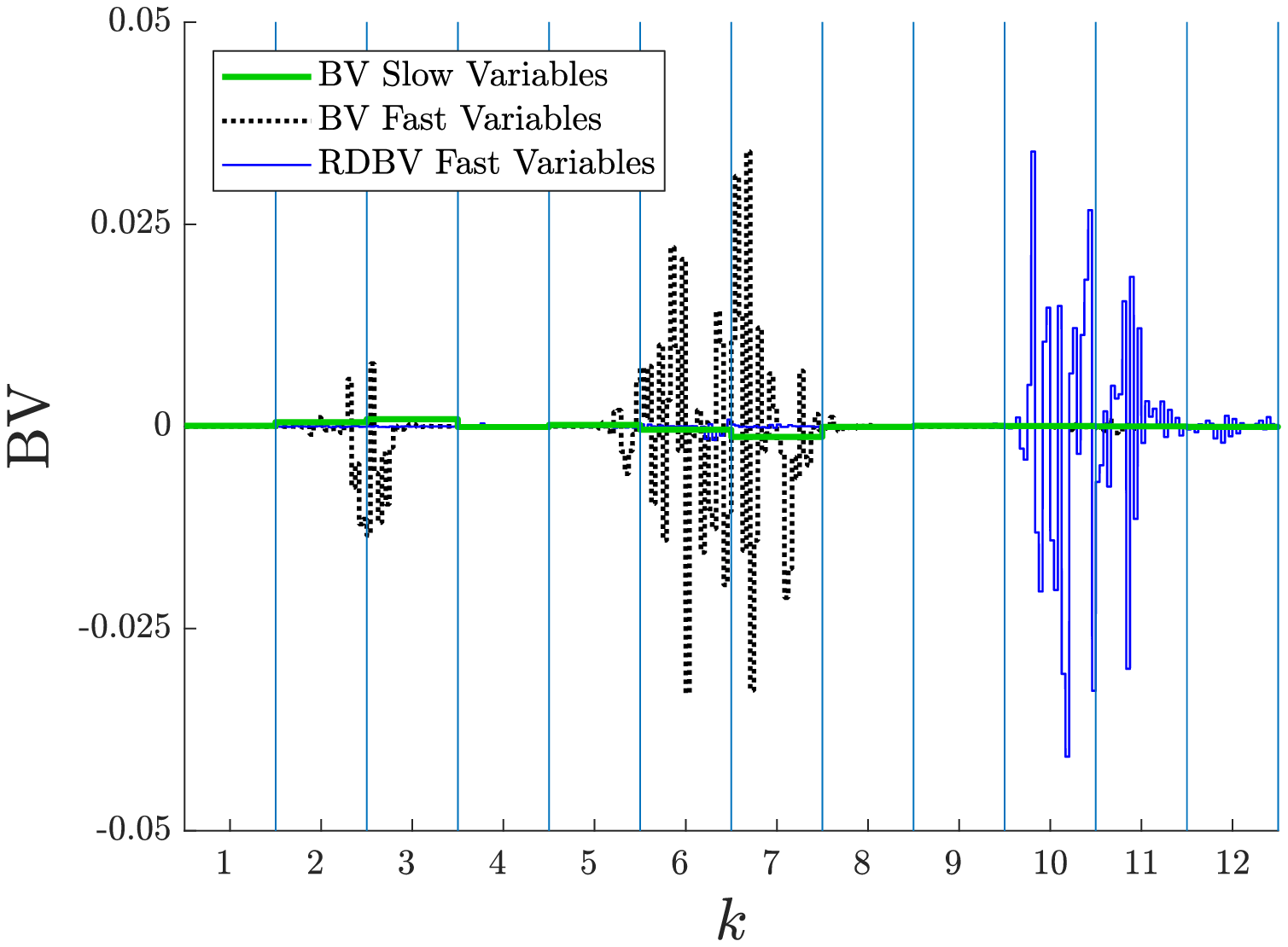}		
			\label{fig:rdbvsnapshot}
		\end{subfigure}
		\caption{
			Stochastically modified bred vectors and associated ${\rm{BV}}$ for the multi-scale L96 model (\ref{e.L96_X})-(\ref{e.L96_Y}) with perturbation size $\delta=0.1$ (cf. Figure~\ref{fig:bvsnapshot}). Shown are the $K=12$ slow components and, for each slow sector, the $J=24$ fast components. (a): SPBV, (b): RDBV.}
	\end{figure}

	An ensemble generated using BVs presents a set of initial conditions which are likely to be propagated into regions of high measure in phase-space. Hence ideally a BV ensemble allows for a sampling of the joint density $\rho(X,Y,\tau)$ at the lead time $\tau$\footnote{We ignore the issue that technically we are dealing in deterministic dynamical systems with measures which are not absolutely continuous with respect to the Lebesgue measure and which are singularly supported only on the attractor.}. In the case when the perturbation size $\delta$ is not sufficiently large to prohibit the collapse of the BV ensemble, the BV was characterised by slow components of small amplitude and by fast components which are localised in the sectors corresponding to the higher amplitude states of the slow variables. Note that BVs in general do not lie on the attractor, but typically relax rapidly onto the attractor along the stable manifold. Each ensemble member therefore is (after some short transient time) drawn from $\rho(X,Y,0)$ (conditioned on fast growth). In multi-scale systems with time-scale separation parameter $\varepsilon=1/c$ the joint density can be approximated (see for example, \citep{GivonEtAl04,PavliotisStuart}) as 
	\begin{align}
	\rho(X,Y,t) = {\hat \rho}(X,t)\rho_\infty(Y|X) + \mathcal{O}(\varepsilon),
	\end{align}	
	where $\rho_\infty(Y|X)$ is the equilibrium probability density function of the fast variables conditioned on the slow variables $X$. The collapse can hence be viewed as having only one single realisation from this fast conditional equilibrium density. Our aim now is to generate additional draws from $\rho_\infty(Y|X)$ from this single realisation. Estimating the fast conditional measure $\rho_\infty(Y|X)$ is computationally too involved. We propose here instead the following simple low-cost method to generate random draws from $\rho_\infty(Y|X)$. Denote by $\boldsymbol{\rrm{H}}$ the projection onto the slow components and by $\boldsymbol{\rrm{h}}$ the projection onto the fast components. We introduce the stochastically-perturbed bred vector (SPBV) of scale $\delta$ by multiplying the fast components of a classical BV $\boldsymbol{b}$ with independent random noise 
	\begin{align}
	\boldsymbol{\rrm{H}}\boldsymbol{b}_{sp} &= \boldsymbol{\rrm{H}}\boldsymbol{b} \nonumber \\
	\boldsymbol{\rrm{h}}\boldsymbol{b}_{sp} &= \delta_{\rm{fast}} \frac{(\Id+ \boldsymbol{\eta})\boldsymbol{\rrm{h}}\boldsymbol{b}}{\| (\Id+ \boldsymbol{\eta})\boldsymbol{\rrm{h}}\boldsymbol{b} \|},
	\label{e.SPBVgen}
	\end{align}	
	with $\boldsymbol{\eta}$ a diagonal $JK\times JK$ matrix with diagonal entries $\eta_{ii} \sim \mathcal N(0,\sigma)$ for sufficiently large $\sigma$. The rescaling size $\delta_{\rm{fast}}$ is the perturbation size of the fast variables only, determined by the requirement that the overall perturbation size of the SPBV $\boldsymbol{b}_{sp}$ is $\delta$. An ensemble of $N$ SPBVs is generated by applying independent stochastic perturbations according to (\ref{e.SPBVgen}) for each of the members of the BV ensemble. The stochastic perturbation is performed only once as a post-processing step when generating initial conditions for a forecast ensemble. The stochastic perturbation essentially acts solely on the dominant components of the BV since we are applying multiplicative noise. It therefore preserves the localised structure associated with the conditioning on the slow $X$ variables. This is illustrated in Figure~\ref{fig:spbvsnapshot} which shows a snapshot of typical SPBV for $\delta=0.1$, together with its parent BV. It is pertinent to mention that the stochastic perturbation causes the SPBV to lie off the attractor. However, since the SPBVs exhibit the same localisation structure as the dynamically consistent BVs, the fast relaxation of the fast variables towards the attractor ensures that after a brief transient SPBVs explore the attractor for fixed slow variables $X$. Hence the SPBVs represent independent draws from the conditional density $\rho_\infty(Y|X)$. 
	The generation of SPBVs is dependent on the variance $\sigma$ of the random perturbation $\boldsymbol\eta$. We will provide numerical evidence in Section~\ref{sec.numerics} (cf. Figure~\ref{fig:L96sigmasens}) that the rescaling of the SPBVs to the size $\delta_{\rm{fast}}$ in (\ref{e.SPBVgen}) causes the properties of SPBVs to statistically converge once the noise strength $\sigma$ is sufficiently large.
	
	
\subsection{{\bf{Random Draw Bred Vectors}}}
	\label{sec.RDBV}
	To highlight the importance of sampling from the conditional distribution $\rho_\infty(Y|X)$ to generate dynamically consistent ensembles we now consider a another stochastic variant of BVs, where we generate random draws from the marginal distribution $\rho_\infty(Y) = \int \rho_\infty(Y|X) dX$ rather than from the conditional distribution $\rho_\infty(Y|X)$. This is achieved by considering again a single classical BV with size $\delta$, and then generating a new ensemble member by replacing its fast components by the fast components of a randomly selected BV from a library of bred vectors with size $\delta$. We coin these random-draw bred vectors (RDBV). In practice, we generate the library on the fly by running $N$ independent simulations simultaneously, that have started from random initial conditions. Each simulation is used to generate one independent BV, where only the fast components are taken and used to produce the RDBVs. As with SPBVs the slow components of an RDBV are left unchanged from the original BV. In RDBVs the fast components are uncorrelated and independent of the slow components which renders RDBVs as almost orthogonal due to the localised character of BVs. An example of an RDBV is shown in Figure~\ref{fig:rdbvsnapshot}. Whereas SPBVs sample the measure of the system locally, conditioned on the slow variables, the dynamically inconsistent perturbations of RDBVs have the potential to drive the dynamics away from the local region of the truth during their relaxation towards the attractor.
	
	
	
\section{{\bf{Numerical results}}}
	\label{sec.numerics}
	We now present results from numerical simulations of the multi-scale L96 system (\ref{e.L96_X})-(\ref{e.L96_Y}), illustrating that SPBVs and RDBVs provide a more diverse and reliable forecast ensemble with improved forecasting skill and furthermore showing that SPBVs are dynamically consistent. We will show in Section~\ref{sec.Dens} that stochastic perturbations of classical BVs increases the diversity of the ensemble. The forecasting skill is investigated in Section~\ref{sec.skill} in terms of the RMS error of the ensemble forecast. The reliability is studied in Section~\ref{sec.reliability} in terms of the RMS error-spread relationship, the continuous ranked probability score and the Talagrand diagram. Sections~\ref{sec.CLV} and \ref{sec.MVL} are concerned with the dynamical consistency of bred vectors; by means of covariant Lyapunov vectors we show that SPBVs project onto the unstable subspace for small to moderate values of $\delta$, and that their temporal evolution is consistent with the true dynamics measured by the mean-variance of their logarithm.
	
\subsection{{\bf{Ensemble dimension}}}
	\label{sec.Dens}
	
\begin{figure}[h]
		\centering
		\includegraphics[width=0.5\linewidth]{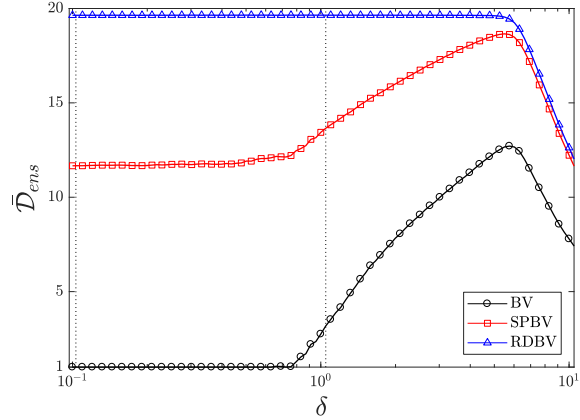}
		\caption{Time-averaged ensemble dimension ${\bar{\mathcal{D}}}_{ens}$ as a function of $\delta$ for each ensemble generation method using a $20$-member ensemble for the multi-scale L96 system (\ref{e.L96_X})--(\ref{e.L96_Y}) with $K=12$ and $J=24$. ${\bar{\mathcal{D}}}_{ens}$ was obtained as an average over $2500$ ensembles. Dashed vertical lines are drawn to delineate values of $\delta=0.103$ and $\delta=1.047$.}
		\label{fig:L96ensdim}
	\end{figure}

	To illustrate the lack of diversity of classical BVs and how stochastically modified BVs improve on diversity, we consider the "ensemble dimension" \citep{BrethertonEtAl99,OczkowskiEtAl05}, also known as the "bred vector dimension" \citep{PatilEtAl01}. The ensemble dimension is a measure for the dimension of the subspace spanned by a set of vectors. For an ensemble of $N$ BV's $\{\boldsymbol{b}^{(n)}(t)\}_{n=1,\dots ,N}$ at a given time, the ensemble dimension is defined as
	\begin{align} 
	\mathcal{D}_{ens}(t) = \frac{\Big( \sum_{n=1}^{N}\sqrt{\mu_n} \Big)^2}{\sum_{n=1}^{N}\mu_n},
	\end{align} 
	where the $\mu_n$'s are the eigenvalues of the $N \times N$ covariance matrix $\mathbf{C}$
	\begin{align} 
	\mathbf{C}_{n,m}(t) = \frac{\boldsymbol{b}^{(n)}(t) [\boldsymbol{b}^{(m)}(t)]^{\mathsf{T}}  }{\| \boldsymbol{b}^{(n)}(t) \|_2 \| \boldsymbol{b}^{(m)}(t) \|_2}.
	\end{align}
	The ensemble dimension takes values between $\mathcal{D}_{ens} = 1$ and $\mathcal{D}_{ens} = {\rm{min}}(N,D)$, where $D=K(J+1)$ is the total dimension of the dynamical system, depending on whether the ensemble members are all aligned or are orthogonal to each other. Figure~\ref{fig:L96ensdim} shows the ensemble dimension ${\bar{\mathcal{D}}}_{ens}$ averaged over $2500$ ensembles as a function of the perturbation size $\delta$ for a 20-member ensemble, for each of the BV ensemble types.\\

	
	Let us first focus on the classical BVs. For a perturbation size $\delta<0.7$, BVs collapse to the local leading Lyapunov vector resulting in an ensemble dimension of ${\bar{\mathcal{D}}}_{ens} = 1$. For $\delta > 0.7$ the perturbations are sufficiently large to allow for nonlinear dynamics to come into effect, resulting in BVs deviating from the local LLV and in an effective increase in ${\bar{\mathcal{D}}}_{ens}$. For $\delta<5.5$ the slow components of the BVs are several orders of magnitude smaller than those associated with the fast variables, with the slow components making up less than $0.2\%$ of the total perturbation size (not shown). Hence, the ensemble dimension is determined by the dynamics in the fast subspace. Once $\delta \approx 4.5$ the fast variables have nonlinearly saturated, meaning that any increase in $\delta$ can only increase the magnitude of the slow components. At $\delta \approx 5.5$, the slow variables make up approximately $K/D = 12/300 = 4\%$ of the total perturbation size (not shown), implying that the slow and fast components are equal in magnitude on average. As $\delta$ continues to increase the slow components of the BVs begin to exceed in size over that of the fast components, rapidly dominating the ensemble dimension. Consequently the ensemble dimension is then only reflecting the dimension of the slow subspace spanned by the perturbations (plus some negligible noise contribution from the fast subspace). Since there are fewer slow variables ($K=12$) than fast variables ($K \times J=288$) the ensemble dimension decreases for  $\delta > 5.5$.
	We indicate in Figure~\ref{fig:L96ensdim} two particular values of the perturbation size $\delta$: $\delta=0.103$ where ${\bar{\mathcal{D}}}_{ens} = 1$ and $\delta=1.047$ where the ensemble dimension of standard BVs has increased to values larger than $1$ and nonlinear effects are active.\\
	
	RDBVs lead to a near maximal ensemble diversity of ${\bar{\mathcal{D}}}_{ens}=19.64$ for the relevant range $\delta<5.5$, since their fast variables are essentially independent random draws with respect to each other. We note that they will not attain the maximal ensemble dimension ${\bar{\mathcal{D}}}_{ens}=N=20$ since the ensemble is not explicitly orthogonalised. Much like for classical BVs, their ensemble dimension decreases for $\delta > 5.5$ for the same reason listed as above since their slow variables are left unchanged from the original BV.\\
	
\begin{figure}[h]
		\centering
		\includegraphics[width=0.5\linewidth]{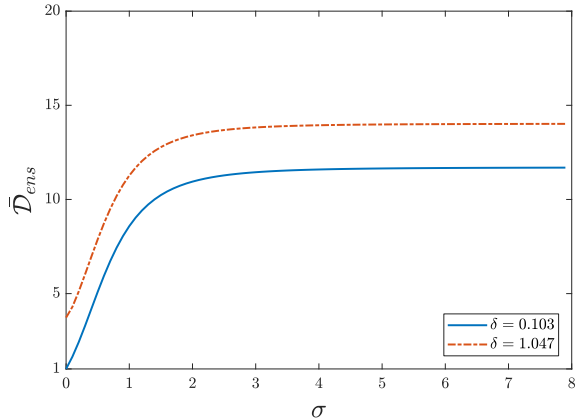}
		\caption{Time-averaged ensemble dimension ${\bar{\mathcal{D}}}_{ens}$ for SPBVs as a function of the noise strength $\sigma$ of the stochastic perturbation for two values of the perturbation size $\delta$. ${\bar{\mathcal{D}}}_{ens}$ was obtained as an average over $2500$ ensembles.}
		\label{fig:L96sigmasens}
	\end{figure}

	SPBVs feature an ensemble dimension ${\bar{\mathcal{D}}}_{ens}\approx 12$ for a perturbation size $\delta<0.7$ \footnote{We remark that observing ${\bar{\mathcal{D}}}_{ens}\approx K$ is accidental. We checked that the ensemble dimension of an SPBV ensemble is related to the number $D_a$ of sites of significant fast activity of the parent BV. Stochastically perturbing BVs with multiplicative noise, as done in (\ref{e.SPBVgen}), then implies that the maximal ensemble dimension is equal to $\rm{min}(N,D_a)=D_a$ for $D_a\approx12 < N=20$.}. Increasing the perturbation size further increases the ensemble dimension which remains significantly larger than that of the classical BVs (recall that each BV in an ensemble is stochastically perturbed). Figure~\ref{fig:L96sigmasens} shows that for sufficiently large values of $\sigma$ the ensemble dimension saturates. The saturation is associated with the rescaling of the perturbation to overall size $\delta$ (cf. (\ref{e.SPBVgen})). This insensitivity of SPBVs to the noise strength $\sigma$ translates to a robustness of other measures presented in the next subsection against changes in $\sigma$. 
	
\subsection{{\bf{Ensemble forecast skill}}}
	\label{sec.skill}
	To study the performance of the various bred vector ensembles, we use here several forecast verification measures to test their predictive ability and their respective uncertainty quantification. We seed an ensemble of bred vectors to be used in a subsequent ensemble forecast around an analysed state obtained during a data assimilation procedure, assuming no model error. We employ here an ensemble Kalman transform filter (ETKF) to provide the analysed state and its associated error covariance $\P_a$. We provide details on the data assimilation procedures used in Appendix~\ref{sec.ETKF}. We consider here the worst case scenario where the classical BVs collapse to a single mode, as well as the case where BVs span a low-dimensional subspace with dimension larger than one but still suffer from under-dispersiveness. We achieve both cases in the L96 system (\ref{e.L96_X})--(\ref{e.L96_Y}) by frequently observing all variables and varying the initial perturbation size of the BVs. 
		To ensure that the ensemble mean of the bred vector ensemble coincides with the analysis, centred pairs of bred vectors with opposite signs are constructed \citep{TothKalnay97,WangEtAl04}. We remark that this is admissible for small values of the perturbation size $\delta$ when bred vectors indeed are vectors; for large values of $\delta$, however, the negative of a bred vector might correspond to an initial condition which is not likely to grow. Alternatively, one may use the simplex method proposed by \cite{Purser96} or may subtract the mean of the rescaled perturbations from all ensemble members \citep{HudsonEtAl13} to centre the ensemble. The perturbation size $\delta$ of bred vectors ideally corresponds to the analysis error, and the perturbation size $\delta$ for the bred vector ideally would be chosen in accordance with the uncertainty of the analysis with $\delta=\sqrt{\Tr\,{\P_a}}$. In many situations, however, this is not sufficient to obtain good forecast skill, as the support of the BV ensemble might not contain the observation. It is common practice to choose $\delta$ larger than the analysis error to achieve better forecasting skill \citep{TothKalnay97,MagnussonEtAl08}. The data assimilation is provided by ETKF with $100$ ensemble members with observations being analysed every $0.005$ time units, matching the breeding cycle. We employ error covariance inflation with a factor of $1.1$. We choose the observational noise to have variances of $5\%$ and $10\%$ of the slow and fast climatic variances of $X$ and $Y$, respectively (cf. Section~\ref{sec.model}). We employ a spin-up period of the ETKF of $25$ time units, before starting the generation of bred vector ensembles for another $25$ time units. We find an average analysis error of $\sqrt{\Tr\,{\P_a}}=0.103$. Ensembles of each bred vector type are generated with $N=20$ members. Each ensemble member is evolved freely, under the same dynamics as the truth, for some lead time $\tau$, at which point the ensemble mean provides the forecast. We report forecast skills for lead times $\tau=1.0$, $\tau=1.5$ and $\tau=2$ time units. New forecasts are produced after each analysis cycle every $0.5$ time units, meaning that there are $100$ BV breeding cycles between forecasts. All metrics are averaged over a total of $M=2500$ forecasts (analysis cycles).\\
	
	To measure the performance of the BV ensembles we consider the root-mean-square error (RMS Error) of the ensemble average with respect to the truth of the slow variables only. With a slight abuse of notation we denote by $X^{(n)}_{k}$ the $k$th slow component of the $n$th ensemble member. Here $k=1,\ldots,K$ and $n=1,\ldots,N$. The ensemble mean is denoted with angular brackets and we have
	\begin{align}
	\langle X_k \rangle = \frac{1}{N}\sum_{n=1}^N X^{(n)}_{k} .
	\end{align}
	
	We define the site-averaged root-mean-square error (RMS Error) between the truth $X^{tr}_k$ and the ensemble average
	\begin{align}
	\mathcal{E}(\tau) = \sqrt{\frac{1}{M}\sum_{m=1}^{M}\frac{1}{K}\sum_{k=1}^{K}
		\| X^{tr}_{k,m}(\tau) - \langle{X}_{k,m}\rangle (\tau) \|^2}
	\end{align}
	as a function of the lead time $\tau$. The index $m$ denotes the realisation with $m=1,\ldots,M$. Similarly, as a measure of the uncertainty of the ensemble average, we consider the site-averaged root-mean-square spread (RMS Spread) 
	\begin{align}
	\mathcal{S}(\tau) = \sqrt{
		\frac{1}{M}\sum_{m=1}^{M}
		\frac{1}{K}\sum_{k=1}^{K}
		\langle
		\| X^{(n)}_{k,m}(\tau) - \langle{X}_{k,m}\rangle (\tau) \|^2\rangle}.
	\end{align}
	
\begin{figure}[h]
		\centering
		\includegraphics[width=0.5\linewidth]{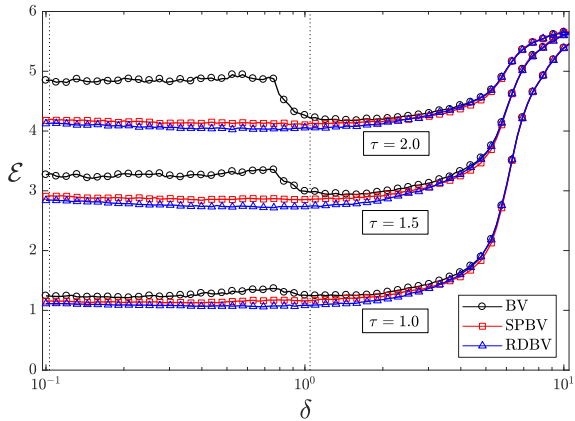}
		\caption{RMS error $\mathcal{E}$ as a function of $\delta$ of each ensemble generation method for three fixed lead times. Dashed vertical lines are drawn to delineate values of $\delta=0.103$ and $\delta=1.047$.}
		\label{fig:L96rmserror}
	\end{figure}

	Figure~\ref{fig:L96rmserror} shows the RMS Error as a function of $\delta$ for three fixed lead times for each ensemble generation method. Classical BVs exhibit the largest RMS errors for $\delta < 0.7$ compared to SPBV and RDBV. This poor performance is due to the fact that for $\delta<0.7$ ensembles of classical BVs suffer ensemble collapse with $\Dens=1$ and the ensemble as a whole eventually diverges from the true state of the system. The diversity of the stochastically modified bred vectors SPBVs and RDBVs causes the ensemble mean to lie closer to the truth with smaller RMS errors. We observe that RDBVs, despite not being dynamically consistent, have the lowest RMS error. Around $\delta \approx 0.7$, when the ensemble dimension of classical BVs increases from a value of $1$, the RMS error of BVs is significantly reduced, and approaches the values of SPBV and RDBV from above upon increasing $\delta$. As $\delta$ is increased further past $\delta\approx 5.5$ the RMS error rapidly rises for all ensemble forecast methods. At this point the slow components of the bred vector begin to dominate in magnitude, resulting in perturbations that are starting much further from the truth. Since our modifications only affect the fast components, all ensemble generation methods then become indistinguishable and yield the same RMS error. Note that for small lead times $\tau=1.0$, the improvement in forecast error of SPBVs and RDBVs over the classical BVs is rather small, suggesting that the ensemble mean has not deviated much from the control forecast.\\
	
\begin{figure}[h]
		\centering
		\includegraphics[width=0.5\linewidth]{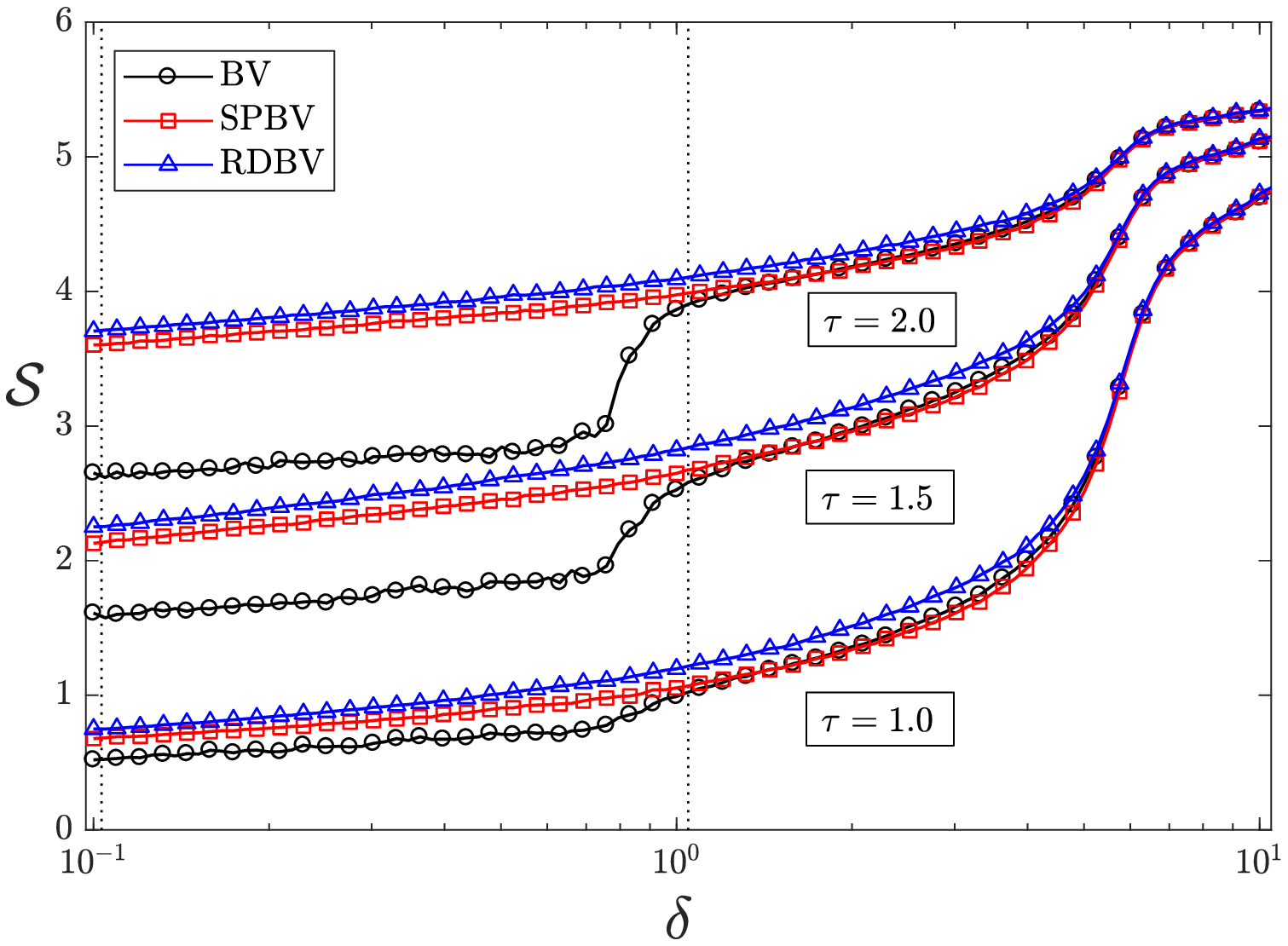}
		\caption{RMS spread $\mathcal{S}$ as a function of $\delta$ of each ensemble generation method for three fixed lead times. Dashed vertical lines are drawn to delineate values of $\delta=0.103$ and $\delta=1.047$.}
		\label{fig:L96rmsspread}
	\end{figure}

	The RMS spread, shown in Figure~\ref{fig:L96rmsspread}, displays a similar behaviour. With $\Dens=1$ the BVs have a very low RMS spread score for $\delta<0.7$. RDBV's spread is slightly higher than the spread of SPBVs; both outperforming classical BVs. Near $\delta \approx 0.7$, the RMS spread of BVs is significantly increased, consistent with the increase of $\Dens$, and approaches the values of SPBV and RDBV from below. As with the RMS error, all methods yield the same RMS spread for $\delta>5.5$.\\
	
	We remark that BVs exhibit a strong dependency of both the RMS error and the RMS spread, when varying the perturbation size $\delta$ from values where the ensemble suffers collapse with $\Dens=1$ to the nonlinear regime where $\Dens>1$. The stochastically perturbed modification SPBVs and RDBVs on the contrary exhibit much less sensitivity of their RMS error and RMS spread when varying the perturbation size. We also note that, for $\delta<0.7$, the RMS error and the RMS spread curves of BVs are less smooth than those of their stochastically modified counterparts. This is because the ensemble average in the RMS quantities is not reducing statistical fluctuations for classical BVs since all BV members are approximately identical in this range.\\ 
	
\begin{figure}[h]
		\centering
		\begin{subfigure}{.33\textwidth}
			\centering
			\caption{}
			\includegraphics[width=0.98\linewidth]{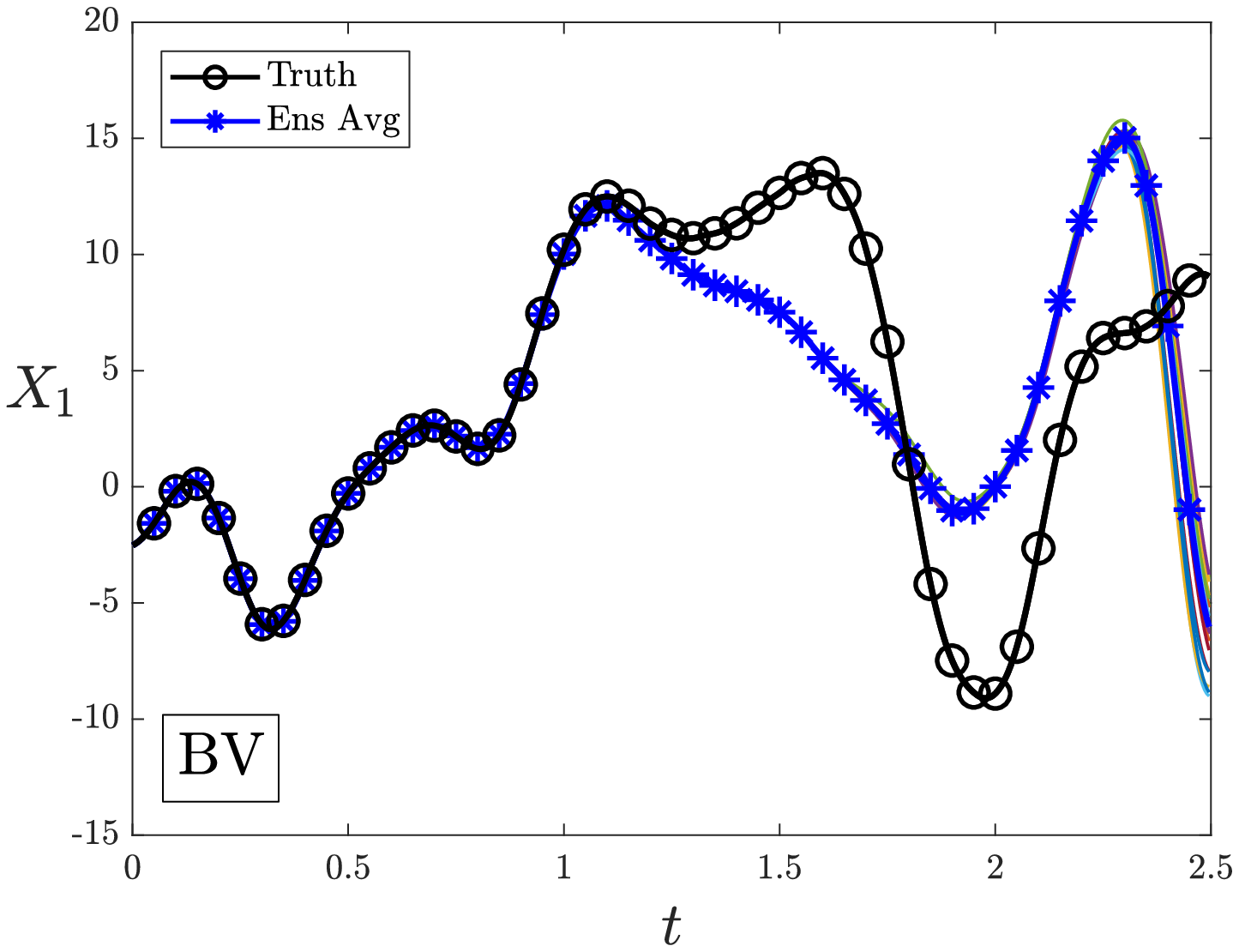}		
			\label{fig:L96BVspag}
		\end{subfigure}%
		\begin{subfigure}{.33\textwidth}
			\centering
			\caption{}
			\includegraphics[width=0.98\linewidth]{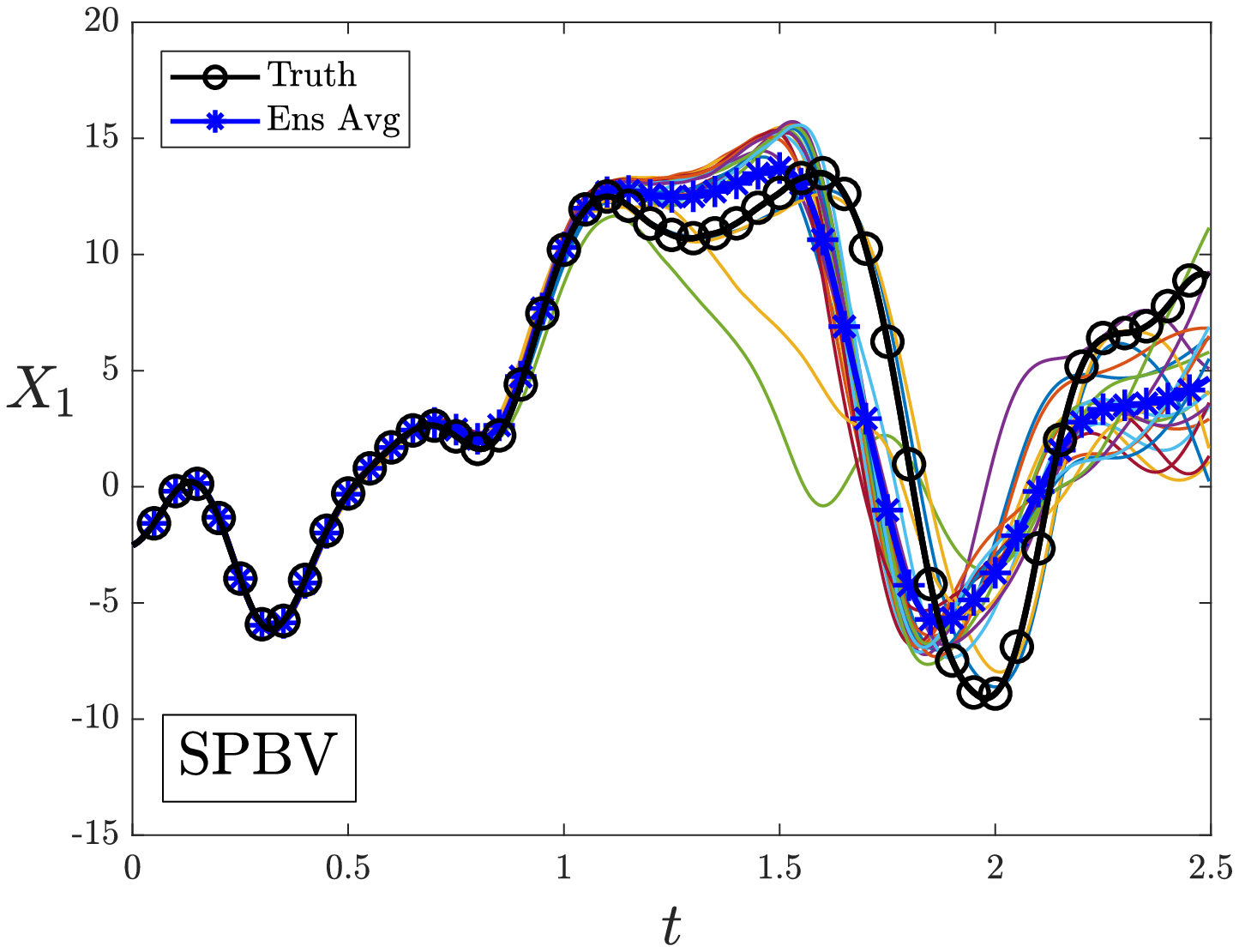}		
			\label{fig:L96SPBVspag}
		\end{subfigure}
		\begin{subfigure}{.33\textwidth}
			\centering
			\caption{}
			\includegraphics[width=0.98\linewidth]{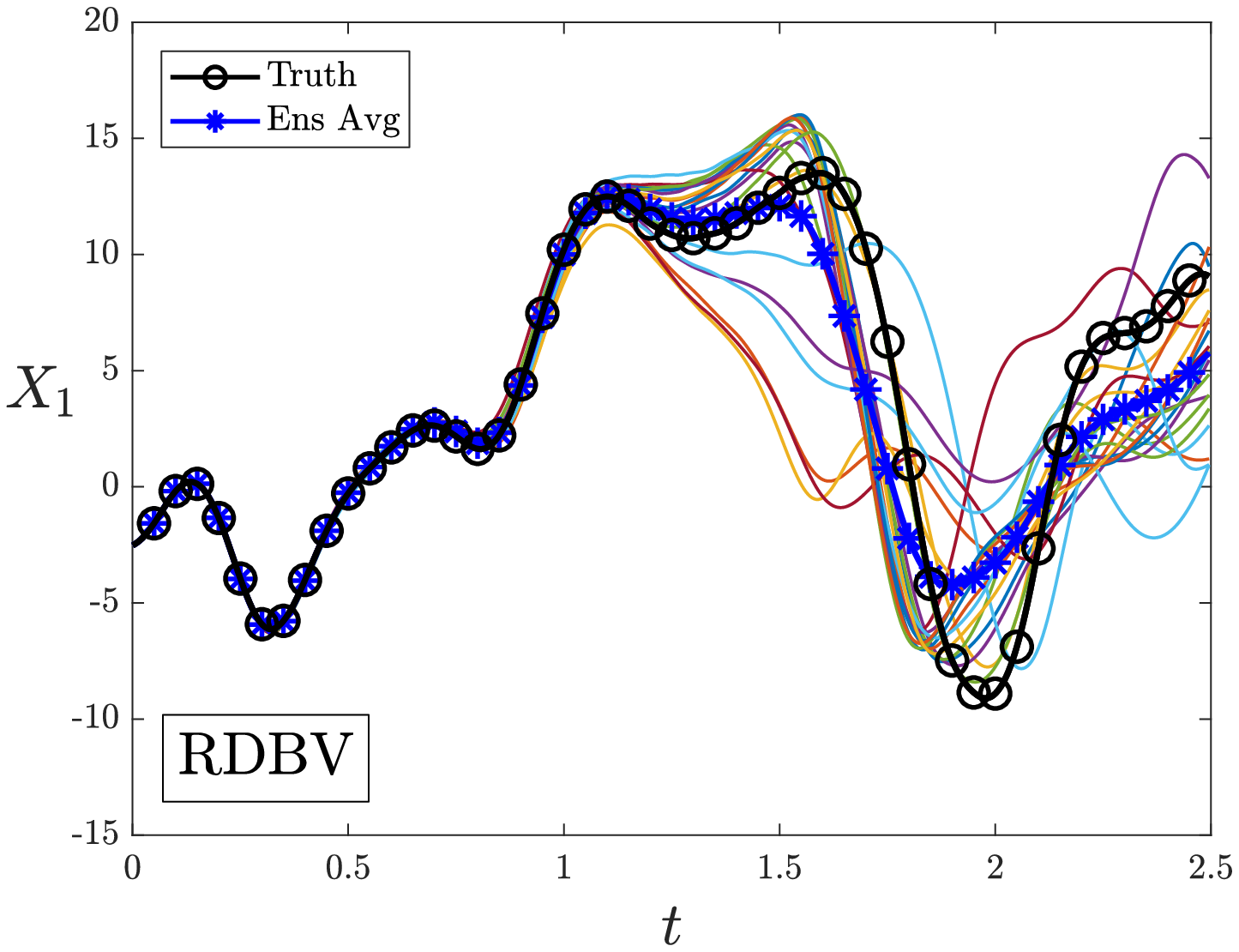}		
			\label{fig:L96RDBVspag}
		\end{subfigure}
		\caption{A typical ensemble forecast for the multi-scale L96 system (\ref{e.L96_X})--(\ref{e.L96_Y}) with $20$ ensemble members. Depicted are the truth, ensemble members and the ensemble average for the slow component $X_{1}$ for (a) BVs, (b) SPBVs and (c) RDBVs. The perturbation size is $\delta=0.1$. For better visibility we only show the BVs and not their negative counterparts.}
		\label{fig:L96spaghettiplots}
	\end{figure}

	
	Figure~\ref{fig:L96spaghettiplots} shows a typical ensemble forecast for the multi-scale L96 system (\ref{e.L96_X})--(\ref{e.L96_Y}) for one slow component $X_1$, with a perturbation size of $\delta = 0.1$. The figure depicts a scenario in which small sub-synoptic perturbations may cause the trajectory to explore several parts of phase-space and acquire synoptic variance. The particular case here allows for two possible synoptic "futures" of the trajectory near lead time $t=1$. We show the truth together with the ensemble mean, as well as all the individual ensemble members. The lack of diversity of BVs is clearly seen to be detrimental and the ensemble collectively diverges from the truth for lead times larger than $1$, causing the poor performance in the RMS error reflected in Figure~\ref{fig:L96rmserror}. The stochastically perturbed bred vectors feature more diversity, allowing some ensemble members to explore the different synoptic "futures" and different parts of the phase space. The spread is particularly large for RDBVs where the fast components are independently drawn, allowing the ensemble to sample more distant parts of phase space. 
	This increased spread leads to less ensemble divergence from the truth and is reflected in the superior RMS error performance of RDBVs (cf. Figure~\ref{fig:L96rmserror}).\\
	
	For the small value of $\delta=0.103$ which corresponds to the average analysis error, all three ensembles have insufficient spread in the sense that the truth has a high likelihood of being outside of the support of the implied probability density function of the ensemble. In the following we therefore consider increased perturbation sizes $\delta=1.047$ for which the RMS error of the BVs is considerably reduced \citep{TothKalnay97,MagnussonEtAl08}. 
		We remark, that if the slow analysed state is too far from the truth of the slow variables, stochastic perturbations of the localised fast components of the BVs will not have any impact on the spread despite increasing the ensemble diversity.
	
	%
	%
	%
	
\subsection{{\bf{Reliability}}}
	\label{sec.reliability}
	In probabilistic forecasting one aims to predict the probability density function at a later time rather than just issuing a forecast of the state. A method which minimises the RMS error of the ensemble forecast mean is not necessarily a method which provides a good probabilistic forecast. Moreover, there are many situations, such as in the case when the probability density function has disjoint support, when the ensemble mean is a bad forecast and is not physically meaningful. We therefore investigate here the reliability of the various bred vector ensembles. In a so called {\it{perfect}} ensemble each ensemble member and the truth are independent draws from the same probability density function $\rho(X,Y)$. In perfect ensembles the ratio between the RMS error of the ensemble mean and the spread of the ensemble approaches $1$ as the ensemble size increases \citep{Wilks,LeutbecherPalmer08}. Ratios smaller or larger than $1$ indicate that the ensemble is either under or over-dispersive, respectively.\\
	
	\begin{figure}[h]
		\centering
		\begin{subfigure}{.48\textwidth}
			\includegraphics[width=0.98\linewidth]{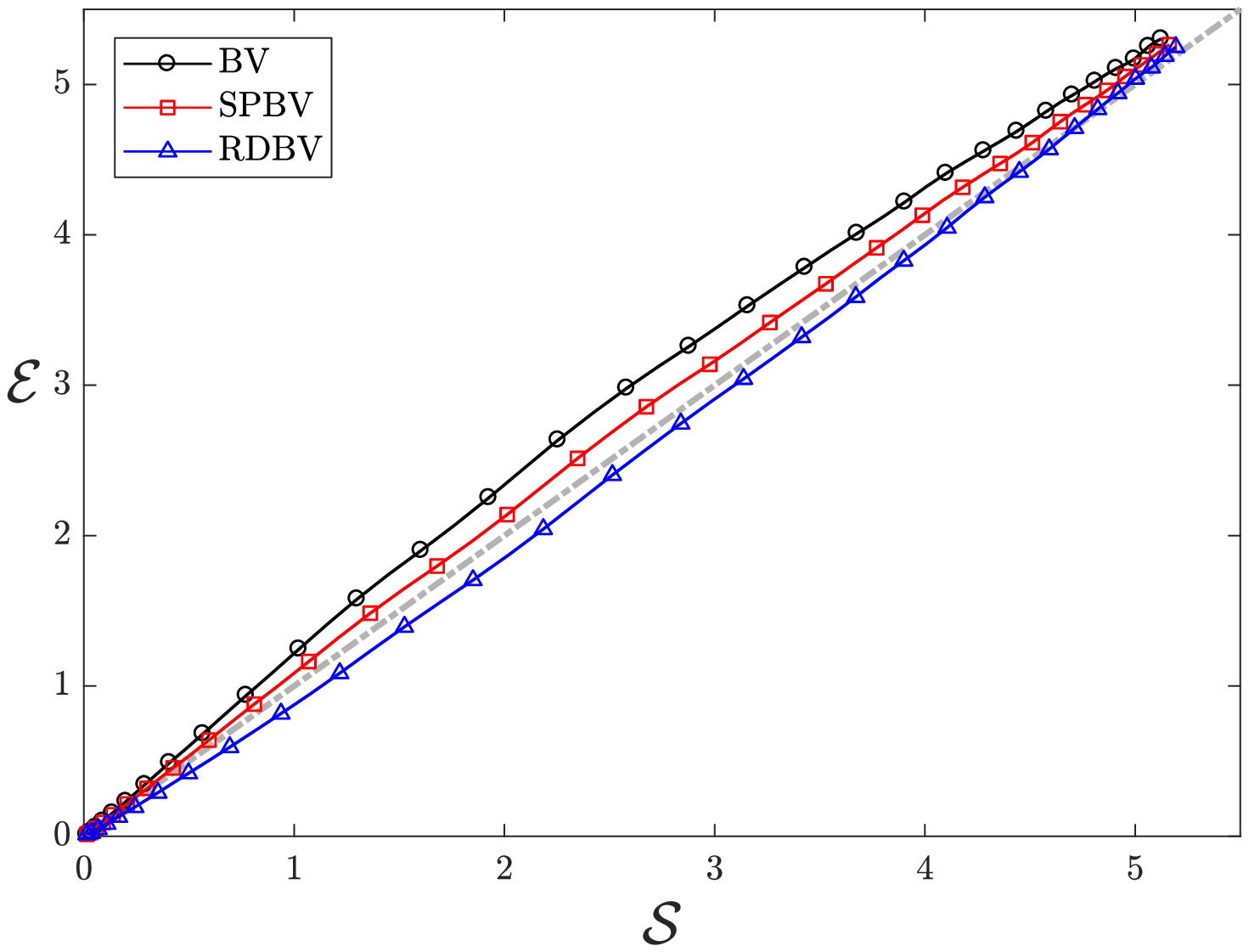}
			\subcaption[]{}
			\label{fig:L96errsprd}
		\end{subfigure}
		\begin{subfigure}{.48\textwidth}
			\includegraphics[width=0.98\linewidth]{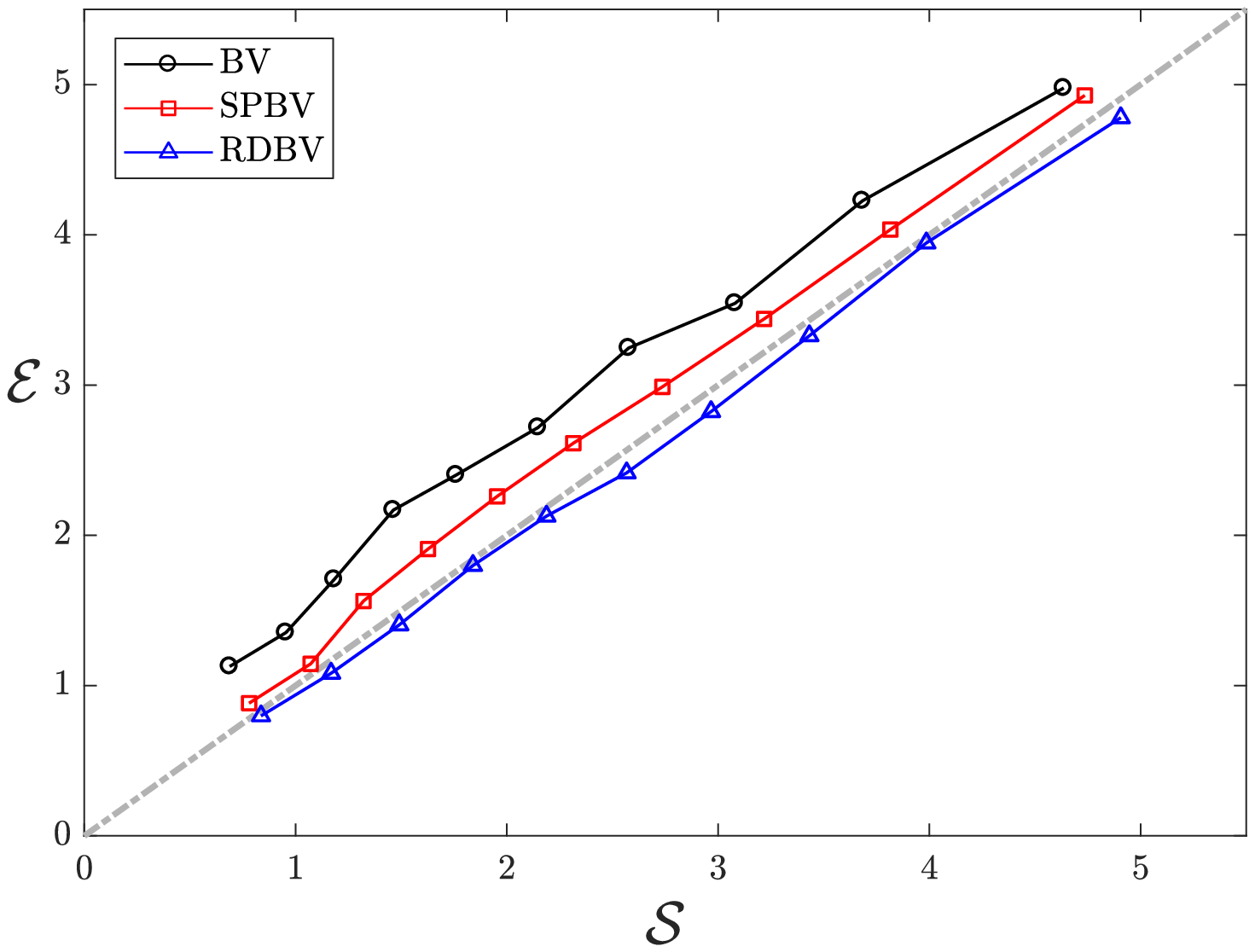}
			\subcaption[]{}
			\label{fig:L96errsprdbin}
		\end{subfigure}
		\caption{(a): Averaged RMS error vs spread for each ensemble generation method parameterised by increasing lead times from $\tau=0$ to $\tau=2.5$ time units for $\delta = 1.047$. (b):  RMS error-spread relationship averaged over $10$ subsamples of increasing RMS spread with lead time $\tau = 1.5$  for $\delta = 1.047$. Each data point represents the average value of each bin and bins increase by $10\%$-percentiles of RMS spread. The grey dot-dashed line indicates a one-to-one ratio of error and spread, indicating a reliable ensemble.}
		\label{fig:L96errsprd_bin}
	\end{figure}
	
	We explore the RMS error versus the RMS spread relationship for the various ensembles for $\delta=1.047$ for which the error dynamics has transitioned from linear to nonlinear dynamics and bred vector ensembles have acquired sufficient diversity to allow for some degree of reliability given the analysis error (cf. Figure~\ref{fig:L96rmserror} and Figure~\ref{fig:L96rmsspread}). Figure~\ref{fig:L96errsprd} shows the RMS error-spread relationship for the various ensembles as a function of the lead time $\tau$, averaged over $2500$ forecasts. Lines lying above/below the one-to-one line indicate under/over-dispersion (error is higher/lower than expected for the amount of spread). BVs are clearly under-dispersive for all lead times. SPBVs have a significantly improved error-spread relationship but still exhibit a certain degree of under-dispersiveness whereas RDBVs are slightly over-dispersive.\\
	
	
	Ideally we would like the RMS error-spread relationship to hold for sufficiently large subsamples of cases conditioned on the predicted spread. In the cases when the predicted ensemble spread is small, this should also be reflected in the error. In a perfect ensemble the spread $\RMSS$ can then be used to predict the standard deviation of the ensemble mean forecast error distribution \citep{WangBishop03,LeutbecherPalmer08}. We therefore stratify the data based on each forecast's individual RMS spread $\RMSS$ for a fixed lead time. This data is placed into $10$ bins of equal size sorted by increasing values of the RMS spread $\RMSS$, so that the first bin contains the forecasts with the 10\%-percentile of $\RMSS$, etc. We then calculate the averaged $\RMSS$ and $\RMSE$ for each bin. These results are displayed in Figure \ref{fig:L96errsprdbin}, again for $\delta=1.047$ at lead time $\tau=1.5$. The results show again that BVs are under-dispersive for all magnitudes of the spread $\RMSS$. 
	SPBVs improve upon classical BVs but still have a small degree of under-dispersiveness, except in the case when the predicted forecast error is small. RDBVs provide the most reliable error spread relationship.\\
	Next, we examine the continuous ranked probability score (\textrm{CRPS}). This is a measure of how well the probabilistic forecast matches the truth combining reliability, uncertainty and resolution of a probabilistic forecast \citep{hersbach00}. The \textrm{CRPS} is defined as
	\begin{align}
	\mathrm{CRPS}(\tau) = \frac{1}{K}\sum_{k=1}^{K}\int_{-\infty}^{\infty} \Big[P(X_k(\tau)) - P^{tr}(X_k(\tau))\Big]^2 \,\mathrm{d}X_k,
	\end{align}
	averaged over each slow variable site $k$ for a particular lead time $\tau$. $P$ and $P^{tr}$ are the cumulative probability distributions of the forecast ensemble and truth, respectively, with
	\begin{align}
	P(X_k) &= \frac{1}{N}\sum_{n=1}^N H\Big(X_k - X^{(n)}_{k}\Big),\\
	P^{tr}(X_k) &= H\Big(X_k-X^{tr}_k\Big),
	\end{align} 
	where $H(x)$ is the Heaviside function with $H(x)=0$ for $x<0$ and $H(x)=1$ otherwise. Here we have made the common choice of estimating the cumulative forecast distribution as a step-function, with steps at the value of each ensemble member $X_k^{(n)}$. \\
	
\begin{figure}[h]
		\centering
		\includegraphics[width=0.5\linewidth]{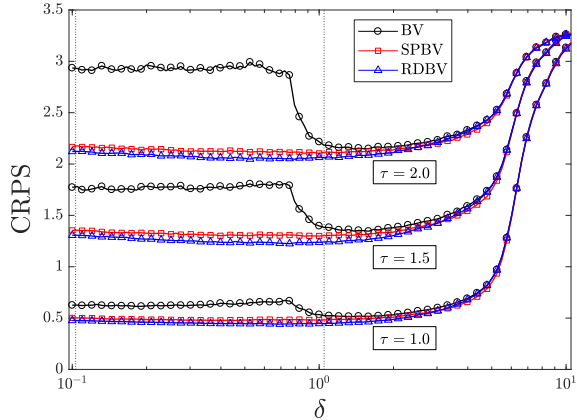}
		\caption{$\rm{CRPS}$ as a function of $\delta$ of each ensemble generation method for three fixed lead times. Dashed vertical lines are drawn to delineate values of $\delta=0.103$ and $\delta=1.047$.}
		\label{fig:L96CRPS}
	\end{figure}

	The features of the \textrm{CRPS} function, depicted in Figure \ref{fig:L96CRPS}, are very similar to that of the RMS error curves shown in Figure~\ref{fig:L96rmserror}. This is not surprising as the \textrm{CRPS} is a generalisation of the mean absolute error \citep{hersbach00}. BVs display the largest \textrm{CRPS} for small values of $\delta$, before sharply decreasing at $\delta \approx 0.7$. SPBVs and RDBVs have very similar values of \textrm{CRPS} across all perturbation sizes, with RDBVs maintaining a slight edge over SPBVs. The most notable difference is that the \textrm{CRPS} punishes the BV ensemble more heavily than the RMS error for $\delta < 0.7$. In particular, for $\tau = 1.0$ BVs exhibit significantly higher values of \textrm{CRPS} compared to SPBVs and RDBVs for $\delta = 0.103$, which is not a feature of the RMS error curve (cf. Figure \ref{fig:L96rmserror}), reflecting the increase in reliability of stochastically perturbed BVs.\\
	
	Another property of a reliable ensemble is that the ensemble spread of the forecast represents the variance of the underlying probability distribution, and that the truth and the ensemble members are statistically indistinguishable random draws from the same probability distribution. 
	This property is conveniently probed in a so called Talagrand or Rank histogram \citep{Anderson96,HamillColucci97,Talagrand99}. To generate a Talagrand histogram, the $N$ ensemble members are sorted at each forecast time and for each variable and used to define a set of $N+1$ bins. We then increment whichever bin the truth falls into at each forecast step to produce a frequency histogram of the truth being in bin $i$. A reliable ensemble then implies that the truth is equally likely to occur in any of those ranked bins leading to a flat histogram.  
	A convex histogram indicates a lack of spread of the ensemble, and a concave diagram indicates an excess of spread of the ensemble \citep{Wilks}. A flat Talagrand diagram does not necessarily imply reliability (see, for example, \citet{Hamill01,Wilks11} for a discussion); however, in the setting of the cyclic Lorenz-96 model (\ref{e.L96_X})-(\ref{e.L96_Y}) with equally weighted ensemble members, flat histograms do imply reliability. \\
	
\begin{figure}[h]
		\centering
		\centering
		\includegraphics[width=0.5\linewidth]{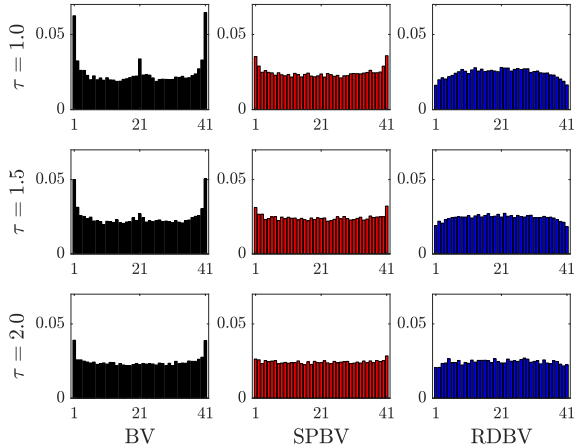}		
		\caption{Talagrand diagrams for $\delta = 1.047$ at lead times $\tau=1.0$, $\tau=1.5$ and $\tau=2.0$ for ensembles of BVs, SPBVs and RDBVs.}
		\label{fig:L96tal}
	\end{figure}

	Figure \ref{fig:L96tal} displays the Talagrand Histogram for each ensemble generation strategy for $\delta=1.047$ for the three lead times $\tau=1$, $\tau=1.5$ and $\tau=2$, averaged over all slow variables. BVs are under-dispersive for all lead times. The peak in the center of the histogram is caused by using $\pm$ BV pairs to center the ensemble around the analysis. RDBVs are shown to be slightly over-dispersive for all lead times, and the degree of over-dispersiveness decreases with increasing lead time $\tau$. SPBVs are shown to be slightly under-dispersive for all lead times, and the degree of under-dispersiveness decreases with increasing lead time $\tau$. Generally, SPBVs and RDBVs form a reliable ensemble for all perturbation sizes for $\tau> 1.5$.
	
\subsection{{\bf{Perturbation dynamics}}}
	\label{sec.evol}
	We now concentrate on the dynamical features of bred vectors and how classical BVs and stochastically perturbed BVs capture the dynamical evolution of perturbations of the full dynamical system (\ref{e.L96_X})-(\ref{e.L96_Y}). Following \citet{PazoEtAl10} we investigate here Lyapunov vectors and the mean-variance of the logarithm (MVL) diagram. 
	
	
\subsubsection{{\bf{Backward and covariant Lyapunov vectors}}}
	\label{sec.CLV}
\begin{figure}[h]
		\centering
		\includegraphics[width=0.5\linewidth]{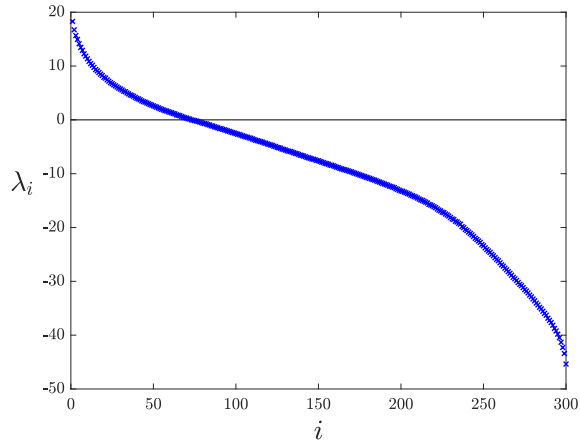}
		\caption{Lyapunov exponent spectrum for the multi-scale L96 model (\ref{e.L96_X})-(\ref{e.L96_Y}) with parameters listed in Table~\ref{tab:L96}.}
		\label{fig:L96lyapexp}
	\end{figure}

	The asymptotic growth of infinitesimal perturbations is captured by Lyapunov vectors. Several types of Lyapunov vectors are commonly used \citep{LegrasVautard96}; for example, one can construct Lyapunov vectors initialised in the asymptotically distant past as so called {\em{backward Lyapunov vectors}}. These backward Lyapunov vectors are generated by solving the linear tangent model of the dynamical system under a Gram-Schmidt orthogonalisation procedure to keep them orthogonal. Each of these backward Lyapunov vectors will evolve, if propagated into the future, to the leading Lyapunov vector and hence backward Lyapunov vectors are not covariant under the tangent dynamics. Covariant Lyapunov vectors, i.e. those for which each Lyapunov vector at time $t$ is propagated to a Lyapunov vector at some later time $t^\prime$ under the linearised dynamics, have been proven to exist under general conditions \citep{Oseledec68}. Contrary to backward Lyapunov vectors, covariant Lyapunov vectors generally do not form an orthogonal basis. \cite{WolfeSamelson07,GinelliEtAl07} designed efficient numerical algorithms to calculate covariant Lyapunov vectors. We remark that, similar to BVs, covariant Lyapunov vectors exhibit a localised "spatial" structure for the L96 system. We use here the algorithm by \cite{GinelliEtAl07} as described in \cite{KuptsovParlitz12} to numerically calculate covariant Lyapunov vectors, with a spin-up period of $250$ time units to ensure convergence of the backward Lyapunov vectors and of $250$ time units to ensure convergence of the expansion coefficients of the covariant Lyapunov vectors. In the following we establish how bred vectors project on backward and onto covariant Lyapunov vectors. For the backward Lyapunov vectors we perform  an orthonormalisation at each time step. Figure~\ref{fig:L96lyapexp} shows the Lyapunov exponents for the multi-scale L96 model (\ref{e.L96_X})-(\ref{e.L96_Y}) with the parameters given in Table~\ref{tab:L96}. There is a total number of $72$ positive Lyapunov exponents. \\
	
\begin{figure}[h]
		\centering
		\begin{subfigure}{.33\textwidth}
			\centering
			\caption{}
			\includegraphics[width=0.98\linewidth]{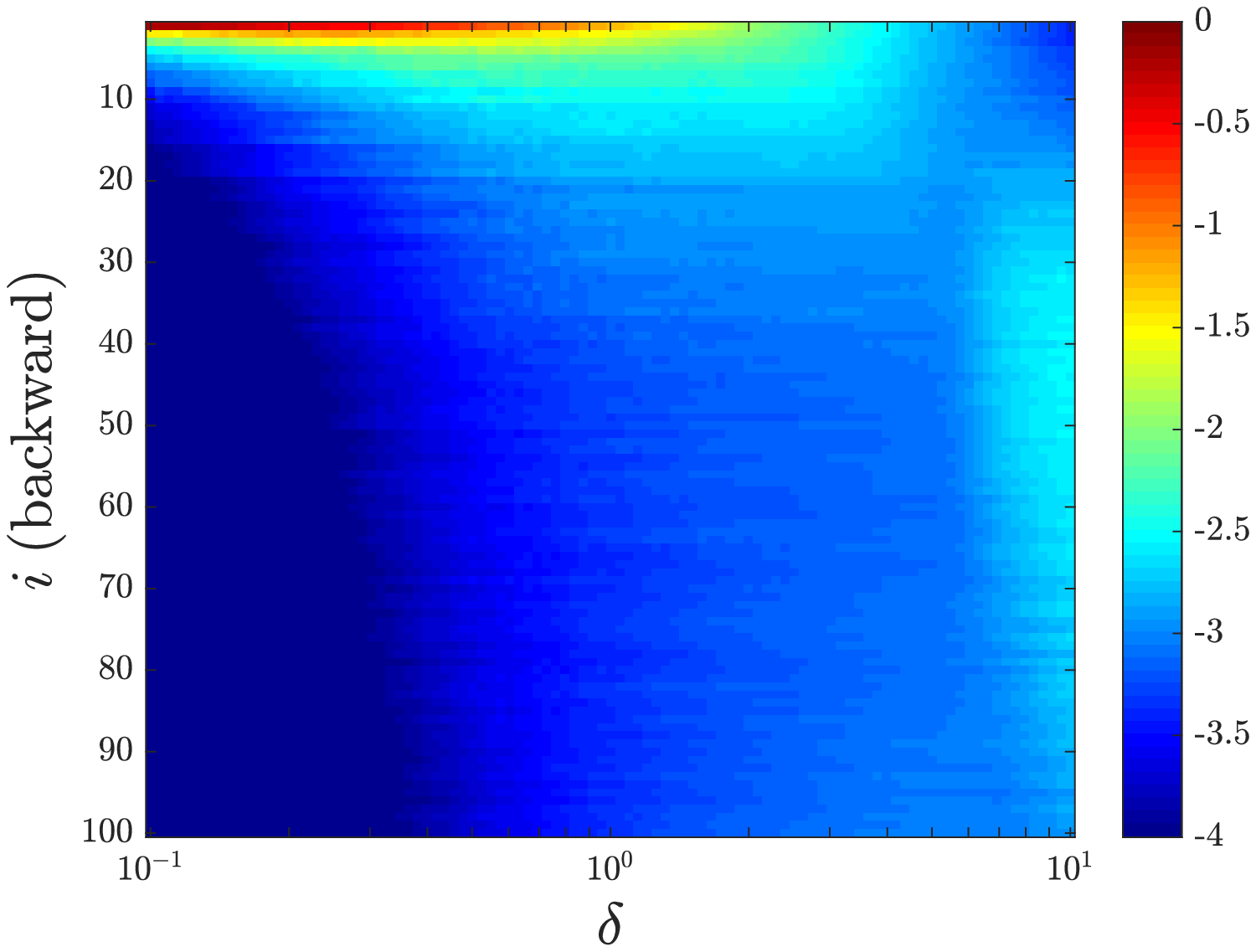}		
			\label{fig:L96BLV_BV}
		\end{subfigure}%
		\begin{subfigure}{.33\textwidth}
			\centering
			\caption{}
			\includegraphics[width=0.98\linewidth]{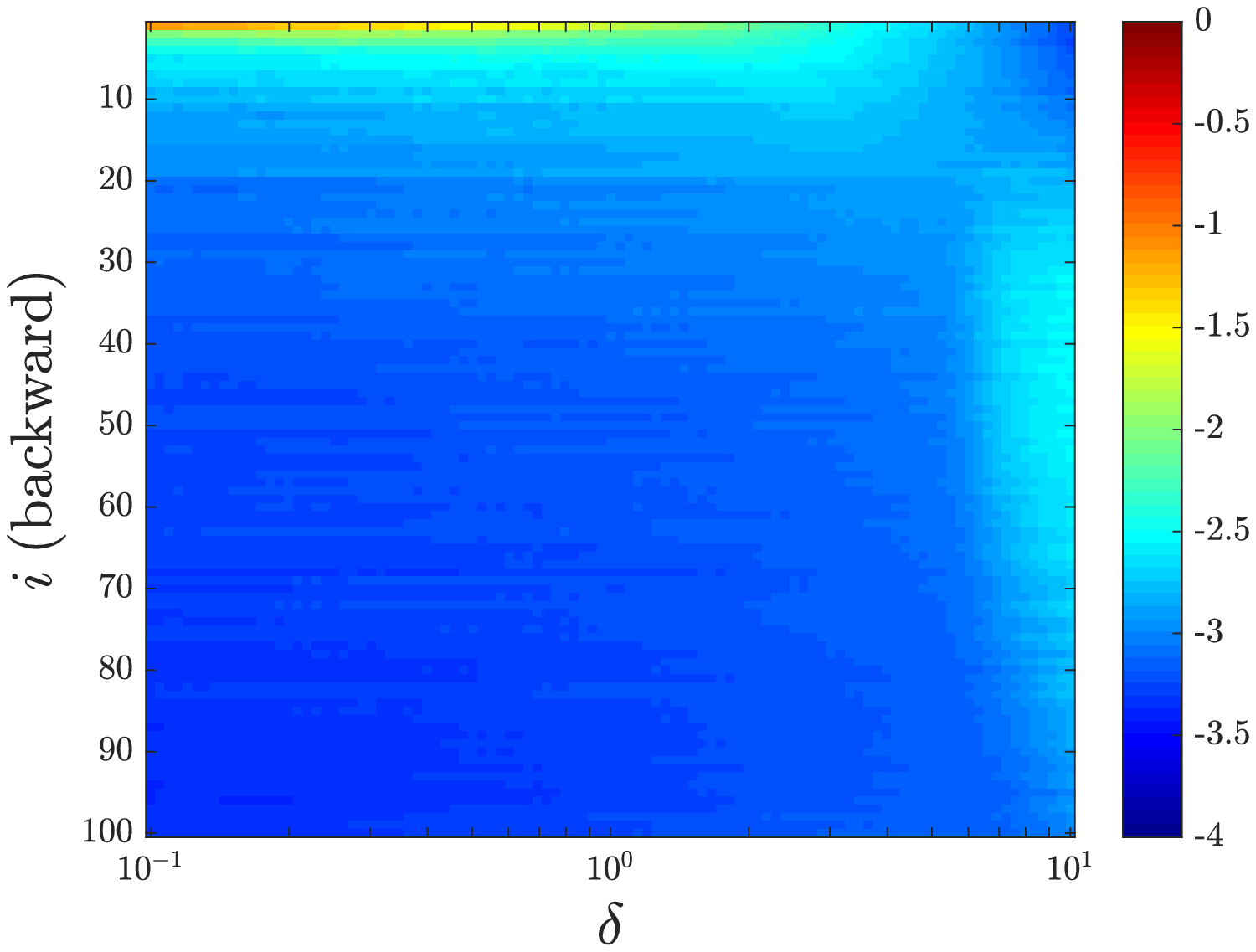}		
			\label{fig:L96BLV_SPBV}
		\end{subfigure}
		\begin{subfigure}{.33\textwidth}
			\centering
			\caption{}
			\includegraphics[width=0.98\linewidth]{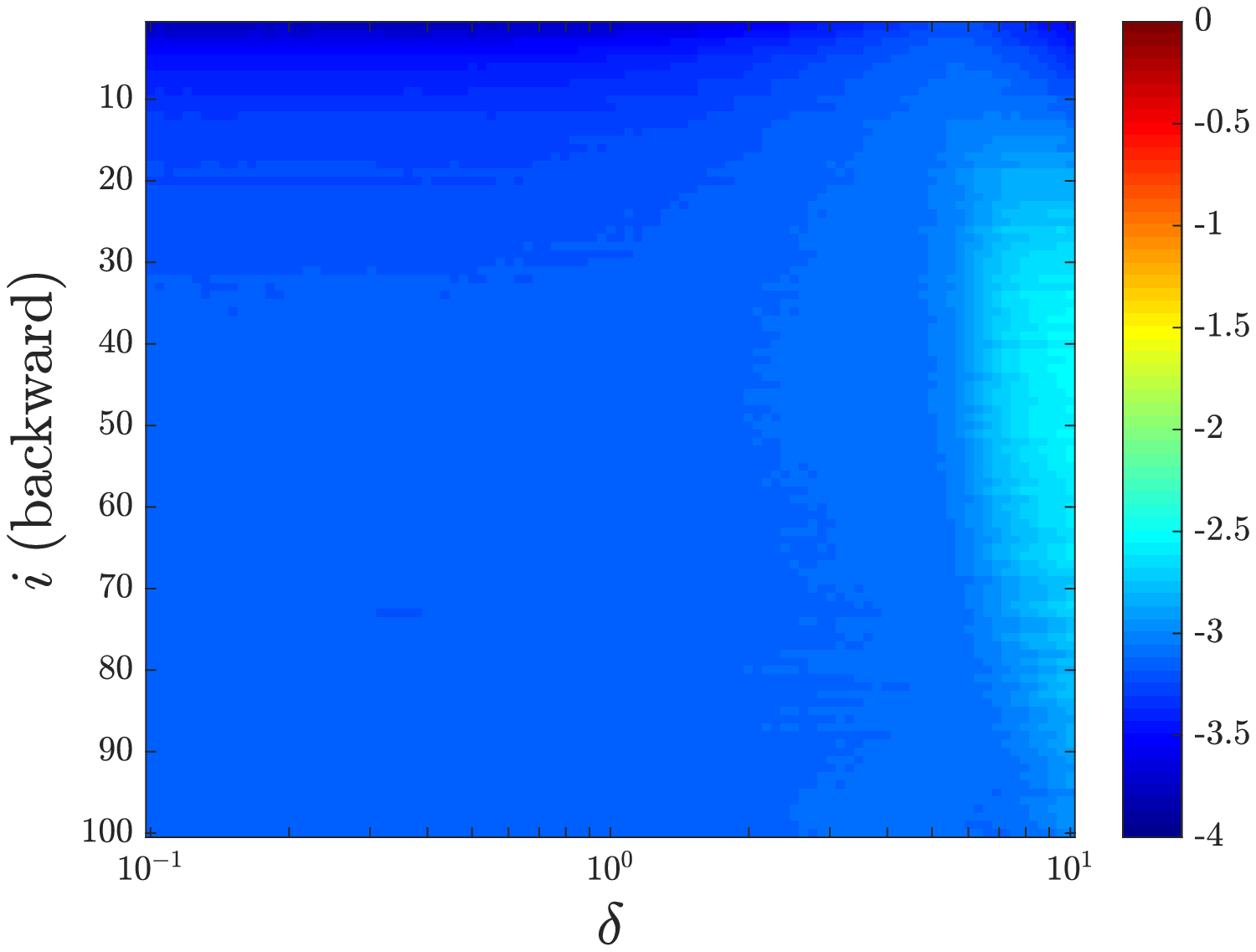}		
			\label{fig:L96BLV_RDBV}
		\end{subfigure}
		\caption{Averaged absolute value projection $\bar\pi_i$ of backward Lyapunov vectors onto (a) BVs, (b) SPBVs and (c) RDBVs. Results are shown on a logarithmic scale.}
		\label{fig:L96BLV}
	\end{figure}

\begin{figure}[h]
		\centering
		\begin{subfigure}{.33\textwidth}
			\centering
			\caption{}
			\includegraphics[width=0.98\linewidth]{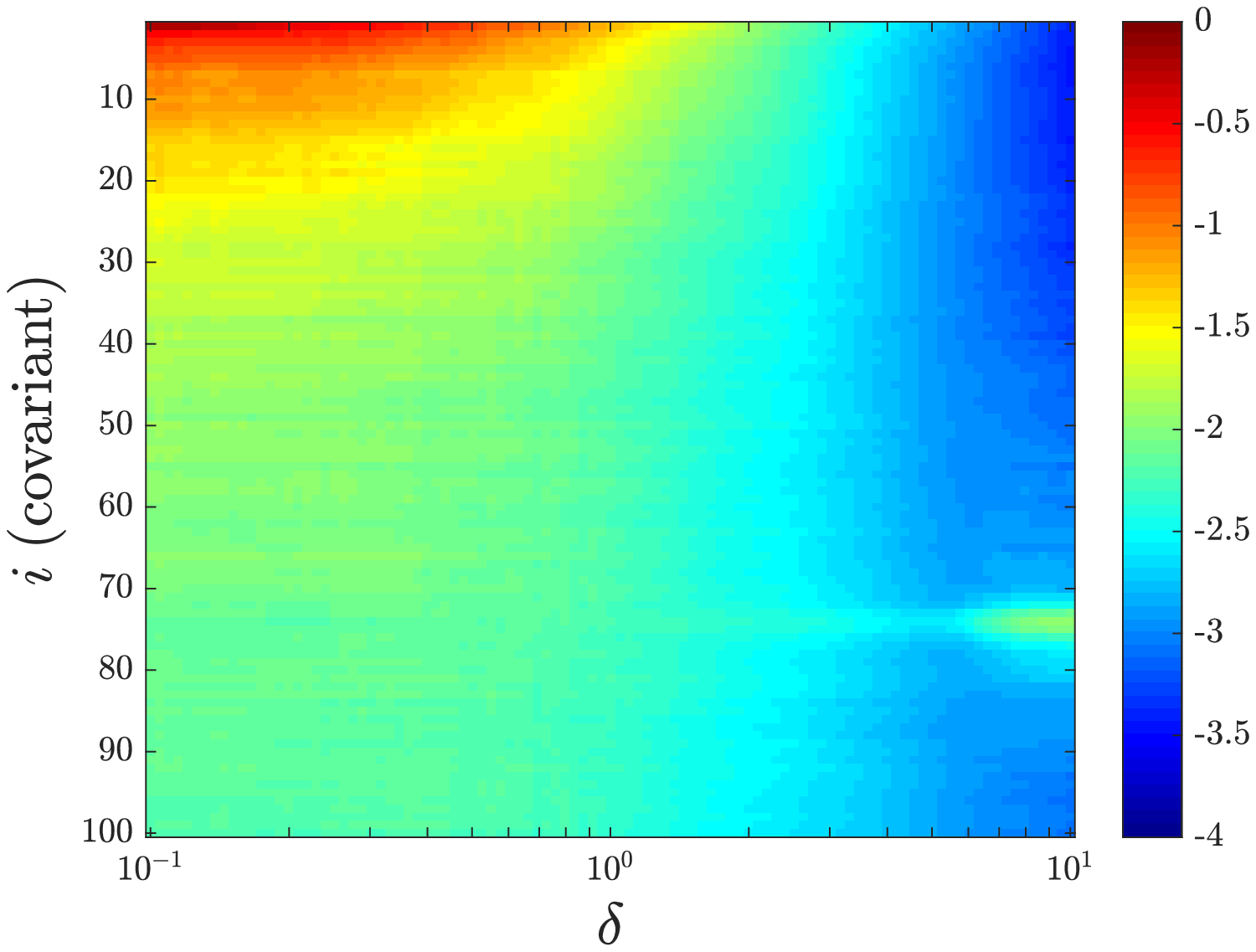}		
			\label{fig:L96CLV_BV}
		\end{subfigure}%
		\begin{subfigure}{.33\textwidth}
			\centering
			\caption{}
			\includegraphics[width=0.98\linewidth]{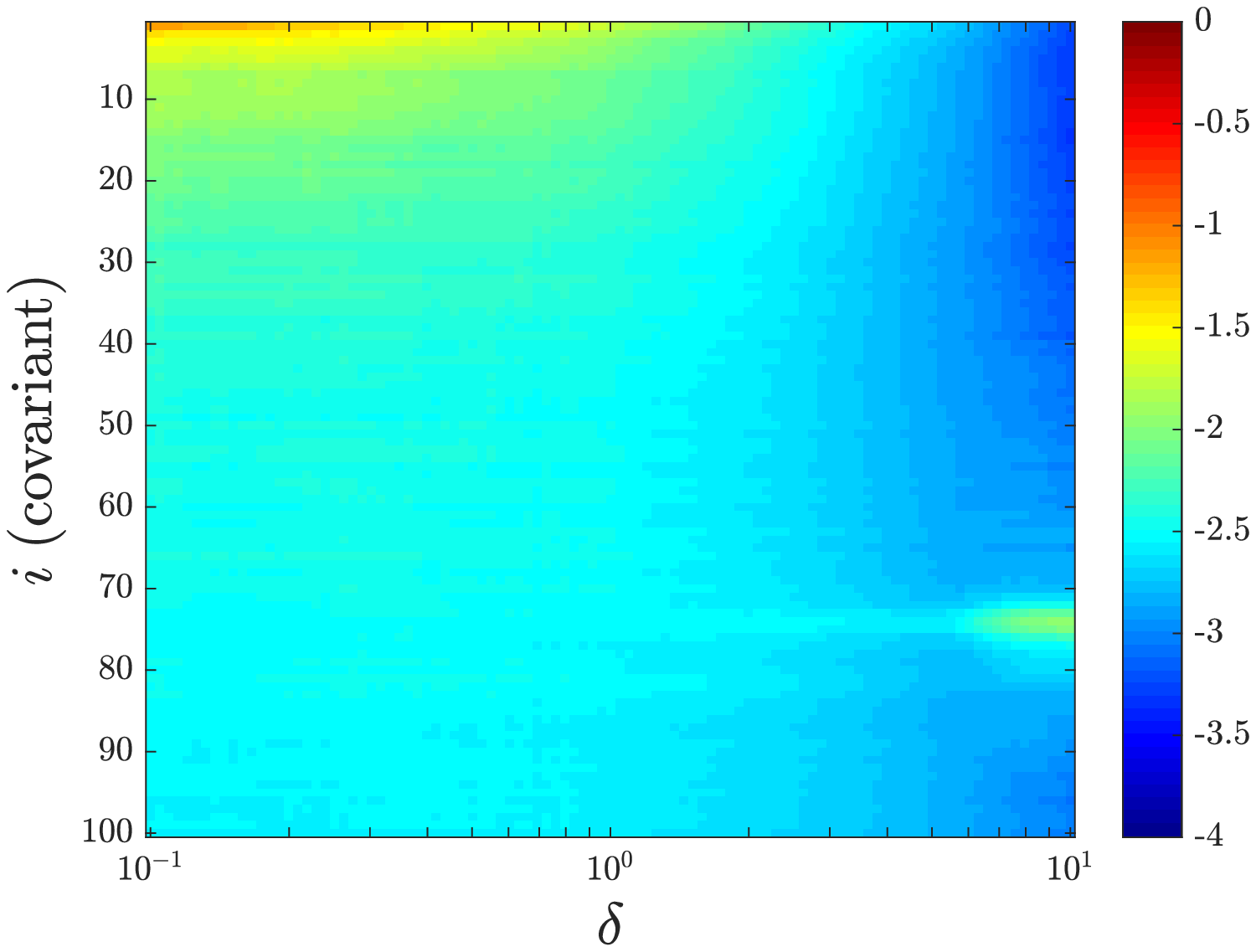}		
			\label{fig:L96CLV_SPBV}
		\end{subfigure}
		\begin{subfigure}{.33\textwidth}
			\centering
			\caption{}
			\includegraphics[width=0.98\linewidth]{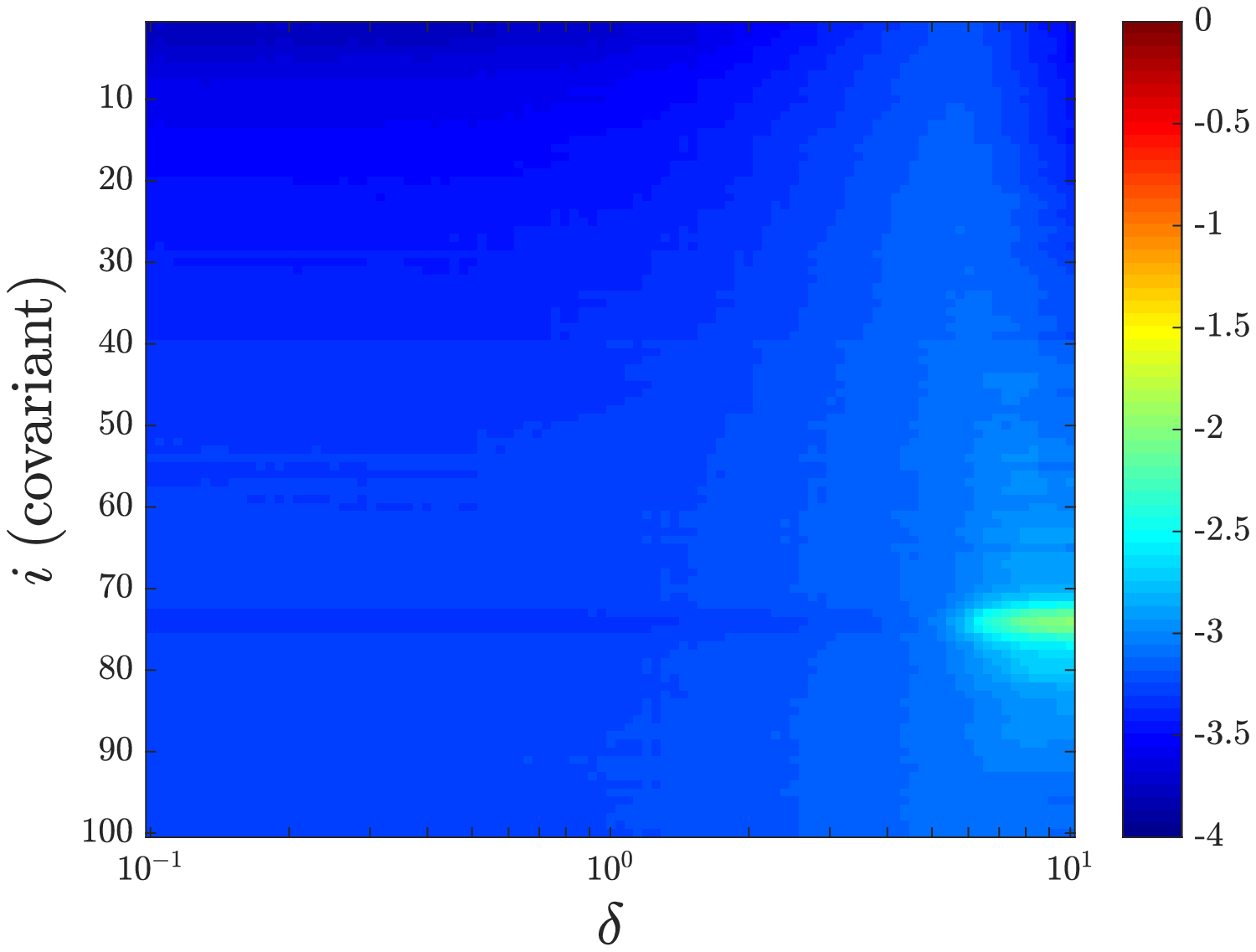}		
			\label{fig:L96CLV_RDBV}
		\end{subfigure}
		\caption{Averaged absolute value projection $\bar\pi_i$ of covariant Lyapunov vectors onto (a) BVs, (b) SPBVs and (c) RDBVs. Results are shown on a logarithmic scale.}
		\label{fig:L96CLV}
	\end{figure}

	To assess the dynamic adaptivity of bred vectors we now study their average projection onto backward and onto covariant Lyapunov vectors. To do so, we normalise the bred vectors and the Lyapunov vectors and introduce the following measure for the degree of projection
	\begin{align}
	\pi^{n}_i(t) = \left\vert 
	\frac{\boldsymbol{b}^{n}(t)}{\| \boldsymbol{b}^{n}(t)\|} \cdot  \frac{\boldsymbol{l}_i(t)}{\| \boldsymbol{l}_{i}(t)\|}
	\right\vert ,
	\end{align}
	where $\boldsymbol{b}^{n}(t)$ denotes the $n$th bred vector ensemble member at time $t$ and $\boldsymbol{l}_i(t)$ denotes the Lyapunov vector corresponding to the $i$th largest Lyapunov exponent at time $t$. We report here on the average $\bar\pi_i$ where we average $\pi^n_i(t)$ over time and over the ensemble members $n$. Hence $\bar \pi_i=1$ corresponds to perfect alignment and $\pi_i=0$ corresponds to (on average) no alignment. Figure~\ref{fig:L96BLV} and \ref{fig:L96CLV} show $\bar\pi_i$ using the first leading $100$ backward and covariant Lyapunov vectors respectively on a logarithmic scale. \\
	
	Let us first focus on the projections of the bred vectors on the backward Lyapunov vectors. It is clearly seen that classical BVs align with the first dominant Lyapunov vectors with indices $i\le3$ for $\delta<0.7$, consistent with the small ensemble dimension $\Dens=1$ in that range, and do not exhibit significant projections onto the orthogonal complement with $i>3$. 
	SPBVs, despite their stochastic perturbation, exhibit a similar collapse on average in the range $i\le3$ but with a lower average projection. This is to be expected since in the range $\delta<0.7$ SPBVs are created from one single collapsed BV and hence they exhibit on average similar projections as their parent BV. RDBVs on the other hand have, as expected, no dominant projection onto any mode since they are composed from BVs at different times. We remark that the projection of BVs and SPBVs for $\delta<0.7$ suggests that indeed BVs can be thought of as linear vectors, spanning (parts of) the unstable subspace. Increasing $\delta$ past $\delta=0.7$ the bred vectors lose their linear character and do not exhibit any significant projection onto the linear Lyapunov vectors. At $\delta \approx  5.5$ when the fast components have saturated, the projection onto the unstable Lyapunov vectors (associated with the fast $Y$ variables) is further remarkably reduced. We remark that for $\delta>5.5$ the bred vectors are dominated by their slow components and all bred vector types (BV, SPBV and RDBV) are essentially indistinguishable as they only vary in their fast components.\\
	
	
	Contrary to backward Lyapunov vectors, covariant Lyapunov vectors do not form an orthogonal basis. Hence, as shown in Figure~\ref{fig:L96CLV}, BVs project onto several unstable covariant Lyapunov vectors (approximately the first $i\le 30$) for $\delta<0.7$. SPBVs, on the other hand, have an average projection of $\bar\pi_i\approx 1$ only for the first three covariant Lyapunov vectors and a lesser but non-trivial contribution onto the vectors with indices $4\le i<10$. Rather than being linked to dynamic properties of SPBVs, this is mainly due to the fact that SPBVs are generated stochastically from a single collapsed BV which causes $\bar \pi_1\approx 1$ due to the ensemble averaging involved in the definition of $\bar\pi_i$. Within an SPBV ensemble, however, individual members may exhibit significant projections onto different covariant Lyapunov vectors (not shown). Both, BVs and SPBVs, have nontrivial projections onto the neutral mode with zero Lyapunov exponent at $i=72$, which is tangential to the flow direction. This is due to the fact that in the L96 model (\ref{e.L96_X})--(\ref{e.L96_Y}) perturbations propagate to the west (i.e. towards decreasing indices $k$ and $j$; cf. Figure\ref{fig:L96heatmap})\footnote{The nonlinear terms in the L96 system (\ref{e.L96_X})--(\ref{e.L96_Y}) can be viewed as a finite-difference discretisation of the advective transport term in geophysical fluid dynamics.} and that BVs and SPBVs exhibit non-trivial activity in the fast and the slow components in the same sectors which are dynamically active. RDBVs, which have unchanged slow components but uncorrelated fast components, naturally show no significant projection onto the linear unstable subspace spanned by the covariant Lyapunov vectors. They only exhibit nontrivial projections onto the neutral mode for large values of $\delta>5.5$ when the slow dynamics is dominant.\\
	
	Our simulations show that BVs exhibit a localisation structure of their fast components which is very similar to that which is exhibited by the first  $20$ covariant Lyapunov vectors, with activity confined to a few well separated spatial regions (cf. Figure~\ref{fig:bvsnapshot} and see also Figure~5 in \cite{HerreraEtAl11}). This is not the case for the first dominant backward Lyapunov vectors which tend to have active fast components in different spatial regions induced by the orthogonality constraint. The confined spatial localisation structure of BVs is inherited by SPBVs, and both BV and SPBV ensembles are dynamically adapted in the sense that their spatial localisation resembles closely that of the dynamical covariant Lyapunov vectors.\\

\subsubsection{{\bf{Evolution of perturbations}}}
	\label{sec.MVL}
	Besides the temporal evolution of error growth, the spatial structure and correlation of perturbations and their evolution encodes important information and has characteristic features which ensemble dynamics should reproduce \citep{LopezEtAl04,GutierrezEtAl08,PazoEtAl10}. In this section we study the free evolution of bred vectors, as done in an ensemble forecast, without rescaling. Rather than studying the evolution of the size of the perturbation $\left\| b_i(t)\right\|$ we now study its logarithm \citep{LopezEtAl04}
	\begin{align}
	h_i(t)=\ln | b_i(t)|,
	\end{align}
	for $i=1,\ldots,D$, where we recall the total dimension of the multi-scale L96 system $D=K(J+1)$, and its spatial mean ${\bar{h}}(t) = \frac{1}{D}\sum_{i=1}^D h_i(t)$. The ensemble averaged spatial mean of the interface of a bred vector is defined as
	\begin{align}
	M(t) = 
	\langle 
	{\bar{h}}(t) 
	\rangle .
	\end{align}
	The variance of the fluctuations around the mean is defined as
	\begin{align}
	V(t) = 
	\langle 
	\frac{1}{D}\sum_{i=1}^D \left( h_i(t) -  {\bar{h}}(t) \right)^2
	\rangle .
	\end{align}
	
\begin{figure}[h]
		\centering
		\begin{subfigure}{.33\textwidth}
			\centering
			\caption{}
			\includegraphics[width=0.98\linewidth]{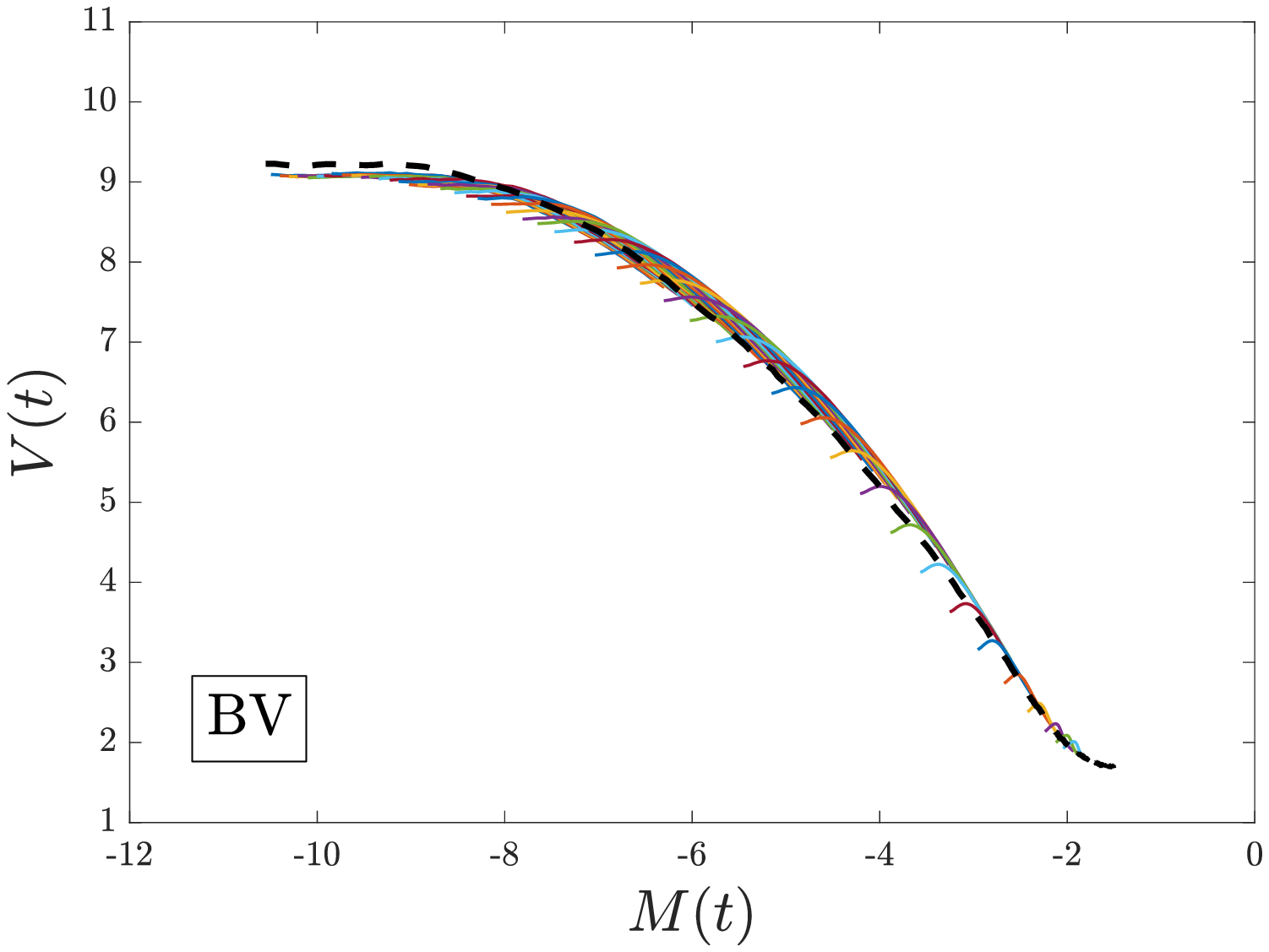}		
			\label{fig:L96mvl_BV}
		\end{subfigure}%
		\begin{subfigure}{.33\textwidth}
			\centering
			\caption{}
			\includegraphics[width=0.98\linewidth]{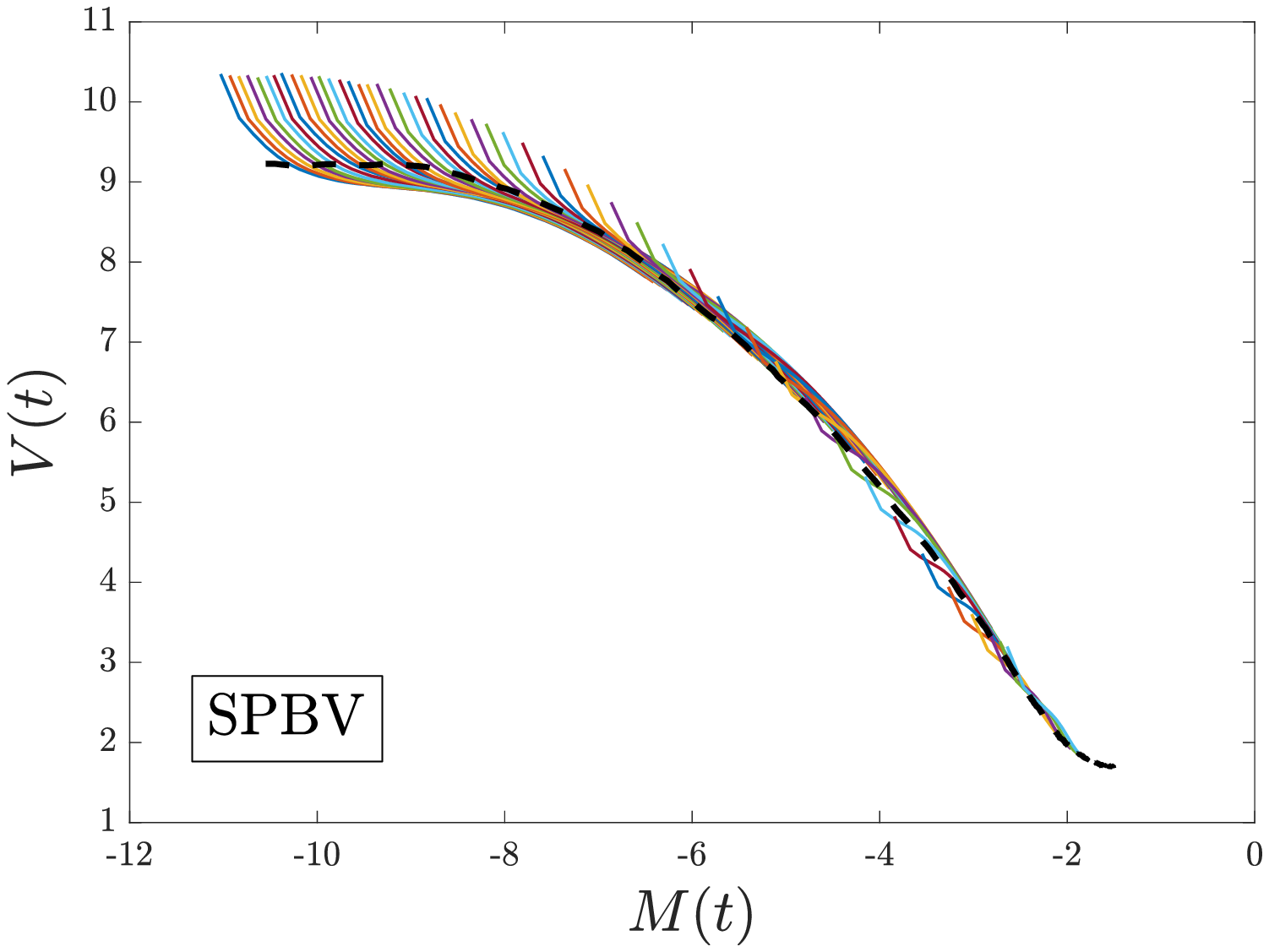}		
			\label{fig:L96mvl_SPBV}
		\end{subfigure}
		\begin{subfigure}{.33\textwidth}
			\centering
			\caption{}
			\includegraphics[width=0.98\linewidth]{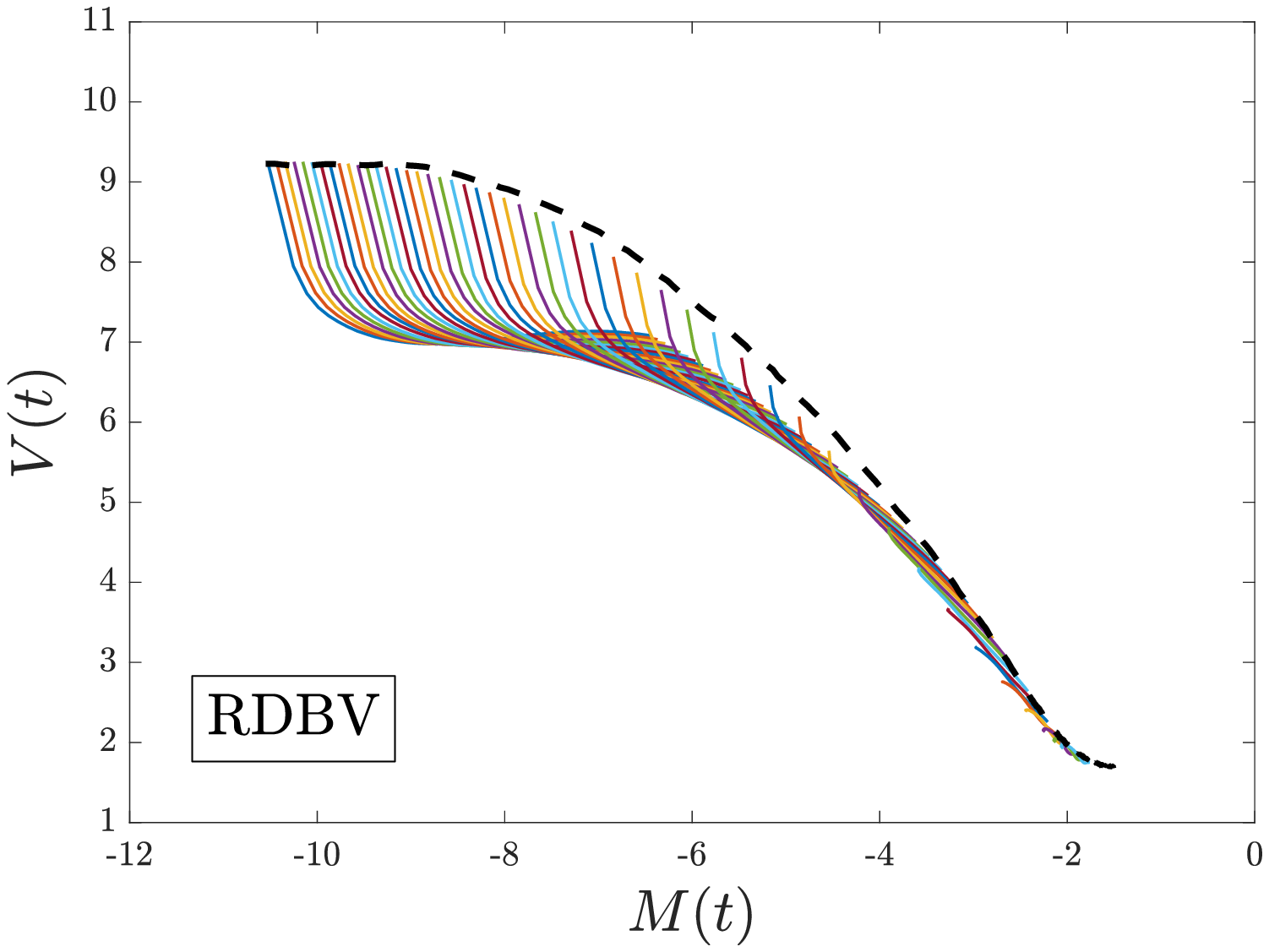}		
			\label{fig:L96mvl_RDBV}
		\end{subfigure}
		\caption{MVL diagram of (a) BVs, (b) SPBVs and (c) RDBVs. Each coloured line represents a different value of $\delta$, increasing from left to right from $\delta=0.1$ to $\delta=10$. 
		The dashed line denotes the MVL curve for the leading Lyapunov vector.}
		\label{fig:L96mvl}
	\end{figure}

	The mean $M(t)$ initially grows linearly in time with the growth rate corresponding to the maximal Lyapunov exponent. In this linear regime the spatial structure is roughly constant with a constant variance $V(t)$. After this initial time, perturbations grow nonlinearly and lose their spatial localisation, i.e. $V(t)$ decreases. In the asymptotic regime $t\to \infty$, the mean saturates to the size of the attractor and the variance decreases until the statistics becomes Gaussian. The mean-variance of the logarithm (MVL) diagram, depicting time traces (as a function of lead time) of $V(t)$ versus $M(t)$, was introduced in \cite{GutierrezEtAl08} to condense the interplay between the temporal mean growth and the spatial growth. MVL diagrams and the characterisation of the spatial structures of bred vectors were used in operational weather prediction models \citep{PrimoEtAl07} and in ensemble prediction systems  \citep{FernandezEtAl09} to compare models. We show in Figure~\ref{fig:L96mvl} the MVL diagram for the multi-scale L96 system (\ref{e.L96_X})--(\ref{e.L96_Y}) where perturbations of classical BVs, SPBVs and RDBVs were taken as initial perturbations, each with $20$ ensemble members, as well as the MVL curve for the leading Lyapunov vector.\\
	
	BVs for perturbation sizes $\delta<0.7$ have an ideal MVL diagram, reproducing the dynamic behaviour of the actual system as characterised by the MVL relationship of the covariant Lyapunov vector. The log-perturbations initially grow linearly in time maintaining a constant variance of $V \approx 9$. Eventually as the trajectory grows further from the truth the mean $M(t)$ increases nonlinearly and the variance of the perturbation declines when $M \approx -8$; curves starting around $M \approx -8$ correspond to perturbations with $\delta\approx 0.7$, where linearity of the perturbation is lost (cf Figures \ref{fig:L96BLV} and \ref{fig:L96CLV}). The MVL diagram for SPBVs clearly reveals that for small perturbation sizes the curves track the reference MVL curve of the leading Lyapunov vector. The initial rapid decline of the variance represents the fast relaxation of the SPBVs towards the attractor with $V=9$; once on the attractor the SPBVs reproduce the error growth behaviour of the leading Lyapunov vector. As the stochastic perturbations of an SPBV ensemble are conditioned on the slow variables, the state towards which the perturbations grow will be close to the actual state of the control analysis forecast. For RDBVs, however, the variance significantly decreases below $V=9$ in its initial phase. This is due to RDBVs being too far off the attractor that their dynamics experiences nonlinear growth of $M$ before developing the spatial localised structure quantified by $V=9$. The RDBVs do not settle on the attractor close to the control analysis forecast but explore large regions of phase space instead. The MVL diagrams show that classical BVs are well adapted to the dynamics of the L96 system. SPBVs are well adapted after a brief transient time needed to relax to the attractor. RDBVs are again, as expected, not dynamically consistent in the sense of the MVL behaviour.
	
\section{{\bf{Discussion and outlook}}}
	\label{sec.summary}
	
	When designing algorithms to generate initial conditions for ensemble forecasting, one would like that the initial ensemble satisfies that they (i) lie on (or at least close to) the attractor, (ii) are reliable, (iii) provide good error skill relationship and (iv) are capable of evolving into areas of large measure \citep{PazoEtAl10}. \cite{PazoEtAl10} performed a detailed comparison between several ensemble methods and argue that CLVs are optimal in the sense of satisfying these four properties; CLVs are, however, computationally very involved. Classical BVs, as we have shown here as well, are dynamically consistent with realistic temporal and spatial error growth evolution, but in the toy model we considered here they lack diversity implying poor forecasting skill and resulting in an under-dispersive ensemble.\\
	
	To mitigate the lack of diversity in classical BV ensembles, we introduced two versions of stochastically modified bred vectors: SPBVs are designed to sample from the equilibrium density of the fast variables conditioned on the slow variables and as such have a theoretical underpinning in the context of multi-scale systems, whereas RDBVs are designed to sample from the marginal equilibrium density of the fast variables and are not conditioned on the slow variables. Both RDBVs and SPBVs significantly improve on forecast skill and on reliability measures compared to classical BVs, with RDBVs outperforming SPBVs with consistently smaller forecast errors. Moreover, whereas the forecast skill of BVs varies significantly when adapting the perturbation size $\delta$ from a regime when they are governed essentially by the linearised dynamics to the nonlinear dynamics, SPBVs' and RDBVs' forecast measures vary smoothly with perturbation size. We showed that SPBVs retain the desirable properties of classical BVs of being dynamically consistent with realistic error growth and realistic evolution of the spatial error structure, but remedy their shortcomings related to their low spread for small perturbation sizes, such as reliability and forecast skill. Furthermore, SPBV ensembles are adapted to the dynamics of the flow in the sense that they exhibit the same localised spatial structure as covariant Lyapunov vectors with very similar temporal and spatial growth.
	
		The superior performance of SPBVs is achieved by the judicious choice of the stochastic perturbation employed to generate them. The stochastic perturbation (\ref{e.SPBVgen}) preserves the localised structure of the parent bred vector. Hence, after a rapid relaxation towards the attractor the initial condition associated with the SPBV will be close to the fiducial trajectory in phase space and the initial condition can be thought of as a random draw from the desired conditional probability measure $\rho(Y|X)$. This causes the spread of an SPBV ensemble to be near optimal in the sense of their RMS error-spread relationship and their Talagrand diagram. RDBVs, on the contrary, do not preserve the local structure of their parent BVs and rather represent an almost orthogonal ensemble. This implies that initial conditions associated with RDBVs are likely to reach the attractor after a brief transient period not near the fiducial trajectory but rather may explore a region in phase space corresponding to different dynamic states. Hence their spread is larger and RDBV ensembles were found to be slightly over-dispersive. Despite being less dynamically consistent and lacking a theoretical justification, the increased diversity of RDBVs allows them to obtain smaller RMS forecast errors when compared to the more theoretically sound SPBVs. We remark that the observed relative ordering of the forecast skill performance of the different bred vector ensembles remains when for each ensemble method the perturbation size $\delta$ is chosen to provide a fixed forecast variance at a fixed lead time $\tau$ (not shown), as suggested, for example, in \citet{TothKalnay97}.\\

	
		The stochastic modifications of bred vectors proposed in this work rely on the localised character of perturbations. This is indeed observed in the multi-scale setting of the L96 system (\ref{e.L96_X})--(\ref{e.L96_Y}). We note that stochastically perturbing localised BVs may not be advantageous and may not generate sufficient spread in situations when the dominant perturbations of the actual dynamics exhibit activity in different sectors than those identified by a bred vector. In this case the multiplicative stochastic perturbation employed for SPBVs would not allow the ensemble to explore the relevant region in phase space. It is planned to extend the ideas of stochastic perturbations to the classical single-scale Lorenz 96 model and explore how they translate to the situation of moderate localisation of perturbations.\\
	
	
	
	From a practical point of view, generating an SPBV ensemble requires the same low computational effort of classical BVs. To generate an ensemble of $N$ SPBVs the propagation of $N$ BVs is required; to generate an ensemble of $N$ RDBVs we would need to evolve $2N$ perturbations (the $N$ BVs as well as $N$ perturbations to sample the fast attractor with different slow initial conditions). Both methods require, of course, far less resources than that which is required for calculating covariant Lyapunov vectors.
	
	\section*{{\bf{Acknowledgements}}}
	BG thanks Diego Paz\'o and Juanma L\'opez for stimulating discussions, pointing us to MVL diagrams and for their hospitality. We would also like to thank the anonymous referees for their constructive and helpful comments. BG acknowledges the support of an Australian Postgraduate Award. GAG acknowledges support from the Australian Research Council, grant DP180101385.
	
	\chapter{Appendix}
	\section{{\bf{Ensemble Kalman filter}}}
	\label{sec.ETKF}
	We briefly describe the ensemble Kalman filter used to obtain the numerical results presented in Section~\ref{sec.numerics}. For more details on data assimilation and ensemble filters the reader is referred to text books such as \cite{Kalnay,Evensen,Simon,ReichCotter,AschBocquetNodet}. In an ensemble Kalman filter (EnKF), proposed by \cite{Evensen94}, an ensemble with $N$ members $\z_n\in \R^D$
	\[
	\Z=\left[ \z_1,\z_2,\dots,\z_N \right]
	\in \mathbb{R}^{D\times N}
	\]
	is propagated by the full nonlinear dynamics ${\dot{\Z}} = {\bf{f}}(\Z)$ with ${\bf{f}}(\Z) =\left[ f(\z_1),f(\z_2),\dots,f(\z_N) \right] \in \mathbb{R}^{D\times N}$. The ensemble is decomposed into its mean 
	\[
	{\bar{\z}} = \frac{1}{N}\sum_{i=1}^N\z_i
	\]
	and its ensemble deviation matrix
	\[
	\Z^\prime=\Z-{\bar{\z}}\e^T ,
	\]
	where $\e=\left[1,\dots,1\right]^T \in \mathbb{R}^{N}$. The ensemble deviation matrix $\Z^\prime$ is used to provide a Monte-Carlo estimate of the forecast covariance matrix
	\[
	\P_f(t)
	= 
	\frac{1}{N-1}\Z^\prime(t)\,\Z^\prime(t)^T
	\in \mathbb{R}^{D\times D} .
	\]
	
	\noindent
	In addition to the forecast ensemble we are also given observations ${\xobs}\in \mathbb{R}^d$ which we express as a perturbed truth with
	\[
	{\xobs}(t_i) = \H \z(t_i) + {\robs} ,
	\]
	where the observation operator $\H:\mathbb{R}^D\to \mathbb{R}^d$ maps from the whole space into observation space, and $\robs \in \mathbb{R}^d$ is i.i.d. observational Gaussian noise with associated error covariance matrix $\Robs$ and zero mean.\\ Given a forecast $\Z_f=\Z(t_i-\epsilon)$ of a chaotic system and its associated forecast error covariance matrix (also known as the \emph{prior}) $\P_f(t_i-\epsilon)$ as well as noisy observations ${\xobs}(t_i)$, data assimilation aims to find the best estimate of the system and updates a forecast into a so-called analysis (also known as the \emph{posterior}). We adopt the convention that evaluation at times $t=t_i - \epsilon$ evaluates a quantity before taking observations $\xobs$, taken at $t=t_i$, into account in the analysis step, and evaluation at times $t=t_i + \epsilon$ evaluates quantities after the analysis step when the observations have been taken into account. 
	
	\noindent
	In the first step of the analysis the forecast mean ${\bar{\z}}_f$ is updated to the analysis mean
	\begin{eqnarray}
	\label{e.zaens}
	{\bar{\z}}_a 
	= 
	{\bar{\z}}_f 
	- \KR\left[\H{\bar{\z}}_f - {\zobs} \right] ,
	\end{eqnarray}
	where the Kalman gain matrix is defined as
	\begin{eqnarray}
	\KR = \P_f \H^T\left( \H \P_f\H^T + \Robs\right)^{-1} .
	\label{e.KGM}
	\end{eqnarray}
	The analysis covariance $\P_a$ is given by 
	\begin{eqnarray}
	\label{e.Pa}
	\P_a = \left(\Id - \KR\H\right) \P_f .
	\end{eqnarray}
	To calculate an ensemble $\Z_a$ which is consistent with the analysis error covariance $\P_a$ in the sense that the ensemble satisfies 
	\[
	\P_a
	= 
	\frac{1}{N-1}\Z_a\,\Z_a^T ,
	\]
	we use the method of deterministic ensemble square root filters which expresses the analysis ensemble as a linear combination of the forecast ensemble. In particular we use the method proposed by \cite{TippettEtAl03,WangEtAl04}, the so called Ensemble Transform Kalman Filter (ETKF). A new forecast $\Z(t_{i+1} -\epsilon)$ is then obtained by propagating $\Z_a(t_i+\epsilon)$ with the full nonlinear dynamics to the next time of observation, where a new analysis cycle will be started.\\ A common problem encountered with ensemble Kalman filters is filter divergence which refers to the problem that in finite ensembles the estimated forecast error covariance $\P_f$ may be too small, potentially prohibiting the analysis to be corrected towards incoming observations, which renders the analysis to be effectively a free running forecast. This underestimation of the error covariance is commonly mitigated by multiplying the error covariance with a so called inflation factor \citep{AndersonAnderson99}. We choose in our simulations $\P_f \to 1.1 \P_f$.\\ 
	
	To seed an ensemble of bred vectors to be used in a subsequent ensemble forecast from an analysis, the perturbation size $\delta$ for the bred vector ideally would be chosen in accordance with the uncertainty of the analysis with $\delta=\sqrt{\Tr\,{\P_a}}$.
	
	%
	

\begin{thebibliography}{68}
\providecommand{\natexlab}[1]{#1}
\providecommand{\url}[1]{\texttt{#1}}
\providecommand{\urlprefix}{URL }
\expandafter\ifx\csname urlstyle\endcsname\relax
  \providecommand{\doi}[1]{doi:\discretionary{}{}{}#1}\else
  \providecommand{\doi}{doi:\discretionary{}{}{}\begingroup
  \urlstyle{rm}\Url}\fi

\bibitem[{Anderson(1996)}]{Anderson96}
Anderson JL. 1996. A method for producing and evaluating probabilistic
  forecasts from ensemble model integrations. \emph{Journal of Climate}
  \textbf{9}(7): 1518--1530.

\bibitem[{Anderson and Anderson(1999)}]{AndersonAnderson99}
Anderson JL, Anderson SL. 1999. {A Monte Carlo implementation of the nonlinear
  filtering problem to produce ensemble assimilations and forecasts}.
  \emph{Monthly Weather Review} \textbf{127}(12): 2741--2758.

\bibitem[{Annan(2004)}]{Annan04}
Annan JD. 2004. On the orthogonality of bred vectors. \emph{Monthly Weather
  Review} \textbf{132}(3): 843--849.

\bibitem[{Asch \emph{et~al.}(2016)Asch, Bocquet and Nodet}]{AschBocquetNodet}
Asch M, Bocquet M, Nodet M. 2016. \emph{Data assimilation}. Society for
  Industrial and Applied Mathematics: Philadelphia, PA.

\bibitem[{Balci \emph{et~al.}(2012)Balci, Mazzucato, Restrepo and
  Sell}]{BalciEtAl12}
Balci N, Mazzucato AL, Restrepo JM, Sell GR. 2012. Ensemble dynamics and bred
  vectors. \emph{Monthly Weather Review} \textbf{140}(7): 2308--2334.

\bibitem[{Bishop \emph{et~al.}(2001)Bishop, Etherton and
  Majumdar}]{BishopEtAl01}
Bishop CH, Etherton BJ, Majumdar SJ. 2001. {Adaptive sampling with the Ensemble
  Transform Kalman Filter. Part I: Theoretical aspects}. \emph{Monthly Weather
  Review} \textbf{129}(3): 420--436.

\bibitem[{Bowler(2006)}]{Bowler06}
Bowler NE. 2006. Comparison of error breeding, singular vectors, random
  perturbations and ensemble {K}alman filter perturbation strategies on a
  simple model. \emph{Tellus A} \textbf{58}(5): 538--548.

\bibitem[{Bretherton \emph{et~al.}(1999)Bretherton, Widmann, Dymnikov, Wallace
  and Blad{\'e}}]{BrethertonEtAl99}
Bretherton CS, Widmann M, Dymnikov VP, Wallace JM, Blad{\'e} I. 1999. The
  effective number of spatial degrees of freedom of a time-varying field.
  \emph{Journal of Climate} \textbf{12}(7): 1990--2009.

\bibitem[{Cai \emph{et~al.}(2003)Cai, Kalnay and Toth}]{CaiEtAl03}
Cai M, Kalnay E, Toth Z. 2003. Bred vectors of the {Z}ebiak-{C}ane model and
  their potential application to {ENSO} predictions. \emph{Journal of Climate}
  \textbf{16}(1): 40--56.

\bibitem[{Cheng \emph{et~al.}(2010)Cheng, Tang, Jackson, Chen and
  Deng}]{ChengEtAl10}
Cheng Y, Tang Y, Jackson P, Chen D, Deng Z. 2010. Ensemble construction and
  verification of the probabilistic {ENSO} prediction in the {LDEO5} model.
  \emph{Journal of Climate} \textbf{23}(20): 5476--5497.

\bibitem[{Corazza \emph{et~al.}(2003)Corazza, Kalnay, Patil, Yang, Morss, Cai,
  Szunyogh, Hunt and Yorke}]{CorazzaEtAl03}
Corazza M, Kalnay E, Patil D, Yang SC, Morss R, Cai M, Szunyogh I, Hunt B,
  Yorke J. 2003. Use of the breeding technique to estimate the structure of the
  analysis "errors of the day". \emph{Nonlinear Processes in Geophysics}
  \textbf{10}(3): 233--243.

\bibitem[{Epstein(1969)}]{Epstein69}
Epstein ES. 1969. Stochastic dynamic prediction. \emph{Tellus} \textbf{21}(6):
  739--759.

\bibitem[{Evensen(1994)}]{Evensen94}
Evensen G. 1994. Sequential data assimilation with a nonlinear
  quasi-geostrophic model using monte carlo methods to forecast error
  statistics. \emph{Journal of Geophysical Research: Oceans} \textbf{99}(C5):
  10\,143--10\,162.

\bibitem[{Evensen(2006)}]{Evensen}
Evensen G. 2006. \emph{{Data Assimilation: The Ensemble {K}alman Filter}}.
  Springer: New York.

\bibitem[{Feng \emph{et~al.}(2016)Feng, Ding, Li and Liu}]{FengEtAl16}
Feng J, Ding R, Li J, Liu D. 2016. Comparison of nonlinear local {L}yapunov
  vectors with bred vectors, random perturbations and ensemble transform
  {K}alman filter strategies in a barotropic model. \emph{Advances in
  Atmospheric Sciences} \textbf{33}(9): 1036--1046.

\bibitem[{Feng \emph{et~al.}(2014)Feng, Ding, Liu and Li}]{FengEtAl14}
Feng J, Ding R, Liu D, Li J. 2014. The application of nonlinear local
  {L}yapunov vectors to ensemble predictions in {L}orenz systems. \emph{Journal
  of the Atmospheric Sciences} \textbf{71}(9): 3554--3567.

\bibitem[{Feng \emph{et~al.}(2018)Feng, Li, Ding and Toth}]{FengEtAl18}
Feng J, Li J, Ding R, Toth Z. 2018. Comparison of nonlinear local {L}yapunov
  vectors and bred vectors in estimating the spatial distribution of error
  growth. \emph{Journal of the Atmospheric Sciences} \textbf{75}(4):
  1073--1087.

\bibitem[{Fern{\'a}ndez \emph{et~al.}(2009)Fern{\'a}ndez, Primo, Cofi{\~{n}}o,
  Guti{\'e}rrez and Rodr{\'i}guez}]{FernandezEtAl09}
Fern{\'a}ndez J, Primo C, Cofi{\~{n}}o AS, Guti{\'e}rrez JM, Rodr{\'i}guez MA.
  2009. {MVL} spatiotemporal analysis for model intercomparison in {EPS}:
  application to the {DEMETER} multi-model ensemble. \emph{Climate Dynamics}
  \textbf{33}(2): 233--243.

\bibitem[{Ginelli \emph{et~al.}(2007)Ginelli, Poggi, Turchi, Chat\'e, Livi and
  Politi}]{GinelliEtAl07}
Ginelli F, Poggi P, Turchi A, Chat\'e H, Livi R, Politi A. 2007. Characterizing
  dynamics with covariant {L}yapunov vectors. \emph{Phys. Rev. Lett.}
  \textbf{99}: 130\,601.

\bibitem[{Givon \emph{et~al.}(2004)Givon, Kupferman and Stuart}]{GivonEtAl04}
Givon D, Kupferman R, Stuart A. 2004. Extracting macroscopic dynamics: {M}odel
  problems and algorithms. \emph{Nonlinearity} \textbf{17}(6): R55--127.

\bibitem[{Greybush \emph{et~al.}(2013)Greybush, Kalnay, Hoffman and
  Wilson}]{GreybushEtAl13}
Greybush SJ, Kalnay E, Hoffman MJ, Wilson RJ. 2013. Identifying {M}artian
  atmospheric instabilities and their physical origins using bred vectors.
  \emph{Quarterly Journal of the Royal Meteorological Society}
  \textbf{139}(672): 639--653.

\bibitem[{Guti\'errez \emph{et~al.}(2008)Guti\'errez, Primo, Rodr\'{\i}guez and
  Fern\'andez}]{GutierrezEtAl08}
Guti\'errez JM, Primo C, Rodr\'{\i}guez MA, Fern\'andez J. 2008. Spatiotemporal
  characterization of {E}nsemble {P}rediction {S}ystems: the mean-variance of
  logarithms ({MVL}) diagram. \emph{Nonlinear Processes in Geophysics}
  \textbf{15}(1): 109--114.

\bibitem[{Hamill(2001)}]{Hamill01}
Hamill TM. 2001. Interpretation of rank histograms for verifying ensemble
  forecasts. \emph{Monthly Weather Review} \textbf{129}(3): 550--560.

\bibitem[{Hamill and Colucci(1997)}]{HamillColucci97}
Hamill TM, Colucci SJ. 1997. Verification of {Eta/RSM} short-range ensemble
  forecasts. \emph{Monthly Weather Review} \textbf{125}(6): 1312--1327.

\bibitem[{Herrera \emph{et~al.}(2010)Herrera, Fern{\'a}ndez, Rodr{\'\i}guez and
  Guti{\'e}rrez}]{HerreraEtAl10}
Herrera S, Fern{\'a}ndez J, Rodr{\'\i}guez M, Guti{\'e}rrez J. 2010.
  Spatio-temporal error growth in the multi-scale {L}orenz'96 model.
  \emph{Nonlinear Processes in Geophysics} \textbf{17}(4): 329.

\bibitem[{Herrera \emph{et~al.}(2011)Herrera, Paz{\'o}, Fern{\'a}ndez and
  Rodr{\'\i}guez}]{HerreraEtAl11}
Herrera S, Paz{\'o} D, Fern{\'a}ndez J, Rodr{\'\i}guez MA. 2011. The role of
  large--scale spatial patterns in the chaotic amplification of perturbations
  in a {L}orenz--96 model. \emph{Tellus A} \textbf{63}(5): 978--990.

\bibitem[{Hersbach(2000)}]{hersbach00}
Hersbach H. 2000. Decomposition of the continuous ranked probability score for
  ensemble prediction systems. \emph{Weather and Forecasting} \textbf{15}(5):
  559--570.

\bibitem[{Hudson \emph{et~al.}(2013)Hudson, Marshall, Yin, Alves and
  Hendon}]{HudsonEtAl13}
Hudson D, Marshall AG, Yin Y, Alves O, Hendon HH. 2013. Improving intraseasonal
  prediction with a new ensemble generation strategy. \emph{Monthly Weather
  Review} \textbf{141}(12): 4429--4449.

\bibitem[{Kalnay(2002)}]{Kalnay}
Kalnay E. 2002. \emph{{Atmospheric Modeling, Data Assimilation and
  Predictability}}. Cambridge University Press: Cambridge.

\bibitem[{Kalnay(2008)}]{KalnayTalk08}
Kalnay E. 2008. Bred vectors: theory and bred vectors: theory and applications
  in operational forecasting. \emph{Six Lectures in Alghero, MSMM08} .

\bibitem[{Keller \emph{et~al.}(2010)Keller, Hense, Kornblueh and
  Rhodin}]{KellerEtAl10}
Keller JD, Hense A, Kornblueh L, Rhodin A. 2010. On the orthogonalization of
  bred vectors. \emph{Weather and Forecasting} \textbf{25}(4): 1219--1234.

\bibitem[{Kuptsov and Parlitz(2012)}]{KuptsovParlitz12}
Kuptsov PV, Parlitz U. 2012. Theory and computation of covariant {L}yapunov
  vectors. \emph{Journal of Nonlinear Science} \textbf{22}(5): 727--762.

\bibitem[{Legras and Vautard(1996)}]{LegrasVautard96}
Legras B, Vautard R. 1996. A quide to {L}yapunov vectors. In: \emph{{S}eminar
  on predictability}, Palmer T\ (ed). ECMWF: Reading, UK, pp. 143--156.

\bibitem[{Leith(1974)}]{Leith74}
Leith CE. 1974. Theoretical skill of {M}onte {C}arlo forecasts. \emph{Monthly
  Weather Review} \textbf{102}(6): 409--418.

\bibitem[{Leutbecher and Palmer(2008)}]{LeutbecherPalmer08}
Leutbecher M, Palmer T. 2008. Ensemble forecasting. \emph{Journal of
  Computational Physics} \textbf{227}(7): 3515 -- 3539. Predicting weather, climate
  and extreme events.

\bibitem[{L\'opez \emph{et~al.}(2004)L\'opez, Primo, Rodr\'{\i}guez and
  Szendro}]{LopezEtAl04}
L\'opez JM, Primo C, Rodr\'{\i}guez MA, Szendro IG. 2004. Scaling properties of
  growing noninfinitesimal perturbations in space-time chaos. \emph{Phys. Rev.
  E} \textbf{70}: 056\,224.

\bibitem[{Lorenz(1963)}]{Lorenz63}
Lorenz EN. 1963. Deterministic nonperiodic flow. \emph{Journal of the
  Atmospheric Sciences} \textbf{20}(2): 130--141.

\bibitem[{Lorenz(1996)}]{Lorenz96}
Lorenz EN. 1996. Predictability: A problem partly solved. In: \emph{Proc.
  Seminar on predictability Vol. 1}, Palmer T\ (ed). ECMWF: Reading, UK, pp.
  1--18.

\bibitem[{Magnusson \emph{et~al.}(2008)Magnusson, Leutbecher and
  K\"all\'en}]{MagnussonEtAl08}
Magnusson L, Leutbecher M, K\"all\'en E. 2008. Comparison between singular
  vectors and breeding vectors as initial perturbations for the {ECMWF}
  ensemble prediction system. \emph{Monthly Weather Review} \textbf{136}(11):
  4092--4104.

\bibitem[{Newman \emph{et~al.}(2004)Newman, Read and Lewis}]{NewmanEtAl04}
Newman CE, Read PL, Lewis SR. 2004. Investigating atmospheric predictability on
  {M}ars using breeding vectors in a general-circulation model. \emph{Quarterly
  Journal of the Royal Meteorological Society} \textbf{130}(603): 2971--2989.

\bibitem[{Norwood \emph{et~al.}(2013)Norwood, Kalnay, Ide, Yang and
  Wolfe}]{NorwoodEtAl13}
Norwood A, Kalnay E, Ide K, Yang SC, Wolfe C. 2013. {L}yapunov, singular and
  bred vectors in a multi-scale system: an empirical exploration of vectors
  related to instabilities. \emph{Journal of Physics A: Mathematical and
  Theoretical} \textbf{46}(25): 254\,021.

\bibitem[{Oczkowski \emph{et~al.}(2005)Oczkowski, Szunyogh and
  Patil}]{OczkowskiEtAl05}
Oczkowski M, Szunyogh I, Patil D. 2005. Mechanisms for the development of
  locally low-dimensional atmospheric dynamics. \emph{Journal of the
  Atmospheric Sciences} \textbf{62}(4): 1135--1156.

\bibitem[{Oseledec(1968)}]{Oseledec68}
Oseledec VI. 1968. A multiplicative ergodic theorem. {C}haracteristic
  {L}japunov, exponents of dynamical systems. \emph{Trudy Moskov. Mat. Ob\v s\v
  c.} \textbf{19}: 179--210.

\bibitem[{O'Kane and Frederiksen(2008)}]{OKaneFrederiksen08}
O'Kane TJ, Frederiksen JS. 2008. A comparison of statistical dynamical and
  ensemble prediction methods during blocking. \emph{Journal of the Atmospheric
  Sciences} \textbf{65}(2): 426--447.

\bibitem[{Palmer(2018)}]{Palmer18}
Palmer T. 2018. The {ECMWF} ensemble prediction system: Looking back (more
  than) 25 years and projecting forward 25 years. \emph{Quarterly Journal of
  the Royal Meteorological Society} \textbf{0}(ja).

\bibitem[{Patil \emph{et~al.}(2001)Patil, Hunt, Kalnay, Yorke and
  Ott}]{PatilEtAl01}
Patil D, Hunt BR, Kalnay E, Yorke JA, Ott E. 2001. Local low dimensionality of
  atmospheric dynamics. \emph{Physical Review Letters} \textbf{86}(26): 5878.

\bibitem[{Pavliotis and Stuart(2008)}]{PavliotisStuart}
Pavliotis GA, Stuart AM. 2008. \emph{{Multiscale {M}ethods: {A}veraging and
  {H}omogenization}}. Springer: New York.

\bibitem[{Paz{\'o} \emph{et~al.}(2013)Paz{\'o}, L{\'o}pez and
  Rodr{\'\i}guez}]{PazoEtAl13}
Paz{\'o} D, L{\'o}pez J, Rodr{\'\i}guez M. 2013. The geometric norm improves
  ensemble forecasting with the breeding method. \emph{Quarterly Journal of the
  Royal Meteorological Society} \textbf{139}(677): 2021--2032.

\bibitem[{Paz{\'o} \emph{et~al.}(2010)Paz{\'o}, Rodr{\'\i}guez and
  L{\'o}pez}]{PazoEtAl10}
Paz{\'o} D, Rodr{\'\i}guez M, L{\'o}pez J. 2010. Spatio-temporal evolution of
  perturbations in ensembles initialized by bred, {L}yapunov and singular
  vectors. \emph{Tellus A} \textbf{62}(1): 10--23.

\bibitem[{Paz{\'o} \emph{et~al.}(2011)Paz{\'o}, Rodr{\'\i}guez and
  L{\'o}pez}]{PazoEtAl11}
Paz{\'o} D, Rodr{\'\i}guez MA, L{\'o}pez JM. 2011. Maximizing the statistical
  diversity of an ensemble of bred vectors by using the geometric norm.
  \emph{Journal of the Atmospheric Sciences} \textbf{68}(7): 1507--1512.

\bibitem[{Pe{\~n}a and Kalnay(2004)}]{PenaEtAl04}
Pe{\~n}a M, Kalnay E. 2004. Separating fast and slow modes in coupled chaotic
  systems. \emph{Nonlinear Processes in Geophysics} \textbf{11}(3): 319--327.

\bibitem[{Primo \emph{et~al.}(2008)Primo, Rodr{\'\i}guez and
  Guti{\'e}rrez}]{PrimoEtAl08}
Primo C, Rodr{\'\i}guez M, Guti{\'e}rrez J. 2008. Logarithmic bred vectors. {A}
  new ensemble method with adjustable spread and calibration time.
  \emph{Journal of Geophysical Research: Atmospheres} \textbf{113}(D5).

\bibitem[{Primo \emph{et~al.}(2007)Primo, Szendro, Rodr\'{\i}guez and
  Guti\'errez}]{PrimoEtAl07}
Primo C, Szendro IG, Rodr\'{\i}guez MA, Guti\'errez JM. 2007. Error growth
  patterns in systems with spatial chaos: {F}rom coupled map lattices to global
  weather models. \emph{Phys. Rev. Lett.} \textbf{98}: 108\,501.

\bibitem[{Purser(1996)}]{Purser96}
Purser RJ. 1996. Arrangement of ensemble in a simplex to produce given first
  and second-moments. \emph{NCEP Internal Report (available from the author at
  Jim.Purser@noaa.gov)} .

\bibitem[{Reich and Cotter(2015)}]{ReichCotter}
Reich S, Cotter C. 2015. \emph{Probabilistic forecasting and {B}ayesian data
  assimilation}. Cambridge University Press, New York.

\bibitem[{Simon(2006)}]{Simon}
Simon DJ. 2006. \emph{{Optimal State Estimation}}. John Wiley \& Sons, Inc.:
  New York.

\bibitem[{Talagrand(1999)}]{Talagrand99}
Talagrand O. 1999. Evaluation of probabilistic prediction systems. In:
  \emph{Workshop proceedings" Workshop on predictability", 20-22 October 1997,
  ECMWF, Reading, UK}.

\bibitem[{Tippett \emph{et~al.}(2003)Tippett, Anderson, Bishop, Hamill and
  Whitaker}]{TippettEtAl03}
Tippett MK, Anderson JL, Bishop CH, Hamill TM, Whitaker JS. 2003. {Ensemble
  square root filters}. \emph{Monthly Weather Review} \textbf{131}(7):
  1485--1490.

\bibitem[{Toth and Kalnay(1993)}]{TothKalnay93}
Toth Z, Kalnay E. 1993. Ensemble forecasting at {NMC}: {T}he generation of
  perturbations. \emph{Bulletin of the American Meteorological Society}
  \textbf{74}(12): 2317--2330.

\bibitem[{Toth and Kalnay(1997)}]{TothKalnay97}
Toth Z, Kalnay E. 1997. Ensemble forecasting at {NCEP} and the breeding method.
  \emph{Monthly Weather Review} \textbf{125}(12): 3297--3319.

\bibitem[{Wang and Bishop(2003)}]{WangBishop03}
Wang X, Bishop CH. 2003. A comparison of breeding and ensemble transform
  {K}alman filter ensemble forecast schemes. \emph{Journal of the Atmospheric
  Sciences} \textbf{60}(9): 1140--1158.

\bibitem[{Wang \emph{et~al.}(2004)Wang, Bishop and Julier}]{WangEtAl04}
Wang X, Bishop CH, Julier SJ. 2004. Which is better, an ensemble of
  positive-negative pairs or a centered spherical simplex ensemble?
  \emph{Monthly Weather Review} \textbf{132}(7): 1590--1605.

\bibitem[{Wei \emph{et~al.}(2008)Wei, Toth, Wobus and Zhu}]{WeiEtAl08}
Wei M, Toth Z, Wobus R, Zhu Y. 2008. Initial perturbations based on the
  ensemble transform ({ET}) technique in the {NCEP} global operational forecast
  system. \emph{Tellus A} \textbf{60}(1): 62--79.

\bibitem[{Wei \emph{et~al.}(2006)Wei, Toth, Wobus, Zhu, Bishop and
  Wang}]{WeiEtAl06}
Wei M, Toth Z, Wobus R, Zhu Y, Bishop C, Wang X. 2006. Ensemble transform
  {K}alman filter-based ensemble perturbations in an operational global
  prediction system at {NCEP}. \emph{Tellus A} \textbf{58}(1): 28--44.

\bibitem[{Wilks(2006)}]{Wilks}
Wilks DS. 2006. \emph{Statistical {M}ethods in the {A}tmospheric {S}ciences}.
  Elsevier: Oxford.

\bibitem[{Wilks(2011)}]{Wilks11}
Wilks DS. 2011. On the reliability of the rank histogram. \emph{Monthly Weather
  Review} \textbf{139}(1): 311--316.

\bibitem[{Wolfe and Samelson(2007)}]{WolfeSamelson07}
Wolfe CL, Samelson RM. 2007. An efficient method for recovering {L}yapunov
  vectors from singular vectors. \emph{Tellus A: Dynamic Meteorology and
  Oceanography} \textbf{59}(3): 355--366.

\bibitem[{Yang \emph{et~al.}(2009)Yang, Keppenne, Rienecker and
  Kalnay}]{YangEtAl09}
Yang SC, Keppenne C, Rienecker M, Kalnay E. 2009. Application of coupled bred
  vectors to seasonal-to-interannual forecasting and ocean data assimilation.
  \emph{Journal of Climate} \textbf{22}(11): 2850--2870.

\end{thebibliography}

\end{document}